\RequirePackage{fix-cm}
\documentclass[natbib,smallextended]{svjour3}       % onecolumn (second format)
\smartqed  % flush right qed marks, e.g. at end of proof
\usepackage{graphicx}
 \journalname{The Astronomy and Astrophysics Review}
\begin{document}

%\title{Investigation tools for tracing the properties of multiple stellar populations in star clusters}
\title{Multiple populations in massive star clusters under the magnifying glass of photometry: Theory and tools}

\subtitle{}

\titlerunning{Cluster multiple populations under the magnifying glass of photometry}        % if too long for running head

\author{Santi Cassisi$^{1,2}$ \and Maurizio Salaris$^3$}

\authorrunning{Cassisi \& Salaris} % if too long for running head

\institute{$^1$INAFÐOsservatorio Astronomico dÕAbruzzo, Via Maggini sn, 64100 Teramo, Italy, \email{santi.cassisi@inaf.it}
\and
$^2$Istituto Nazionale di Fisica Nucleare (INFN) - Sezione di Pisa, Universit‡ di Pisa, Largo Pontecorvo 3, 56127 Pisa, Italy
\and
$^3$Astrophysics Research Institute, Liverpool John Moores University, IC2, Liverpool Science Park, 146 Brownlow Hill, Liverpool, L3 5RF, UK}

\date{Received: date / Accepted: date}

\maketitle

\begin{abstract}
The existence of star-to-star light-element abundance variations in massive Galactic
and extragalactic star clusters has fairly recently superseded the traditional paradigm of individual 
clusters hosting stars with the same age, and uniform chemical composition.
Several scenarios have been put forward to explain the origin of this multiple stellar population
phenomenon, but so far all have failed to reproduce the whole range of key observations.

Complementary to high-resolution spectroscopy, which has first revealed and characterized chemically
the presence of multiple populations in Galactic globular clusters,
photometry has been instrumental in investigating 
this phenomenon in much larger samples of stars --adding 
 a number of crucial observational constraints and correlations
with global cluster properties-- and in the discovery and characterization of multiple populations also in
Magellanic Clouds' intermediate-age clusters.

The purpose of this review is to present the theoretical underpinning and application of the photometric
techniques devised to identify and study multiple populations in resolved star clusters.
These methods have played and continue to play
a crucial role in advancing our knowledge of the cluster multiple population phenomenon, 
and promise to extend the scope of these investigations to resolved clusters even beyond the Local Group, with 
the launch of the James Webb Space Telescope.

\keywords{Globular clusters: general \and Infrared: stars \and opacity \and Stars: evolution \and Stars: imaging
  \and Ultraviolet: stars}

\end{abstract}

\section{Introduction}
\label{intro}

Globular clusters (GCs) are widely employed as tracers of Galaxy evolution, due to their old
ages (on the order of 10~Gyr) which imply a formation redshift $z \sim$2 or higher,
around the peak of the cosmic star formation \citep[see, e.g.,][]{md14}.
Their survival until today allows us to treat GCs as fossil records, whose stars preserve
the chemical and dynamical properties of their origin.

The formation of GCs is however 
still an open problem \citep[see, e.g.,][for a recent review]{forbes},
made even more complex by the discovery that they do not host simple (single-age, single chemical composition) stellar 
populations, as generally believed. It is since about 40 years that 
variations of the initial chemical abundances of some light elements in individual Milky Way GCs have been revealed, 
\citep[see, e.g.,][]{cohen},
however only the much more recent advent of high-resolution multi-object spectrographs has firmly established this
result \citep[see, e.g.,][and references therein]{carretta09a, carretta09b, gcb:12}.

In addition to direct spectroscopic measurements, intracluster abundance variations have been revealed also
through photometry, due to their effect on stellar effective temperatures, luminosities,
and spectral energy distributions
\citep[see, e.g.,][]{salaris:06, M4UBV, yong:08, sswc:11, cmpsf:13, dale16, mucc16, m17,
  dale18, s19}. 
The use of appropriate colours and colour combinations (denoted as \textit{colour indices} or \textit{pseudocolours}) 
has indeed allowed us to greatly enlarge the sample of clusters investigated, the sample of stars surveyed
in individual clusters, and the range of evolutionary phases (including the main sequence, 
typically too faint to be investigated spectroscopically
with current observational facilities) where chemical abundance variations have been detected \citep[see, e.g.,][]{milone:12a, sumo, p15, m17, nieder17}.
%Very specifically, photometric studies have for example clearly demonstrated the existence of He abundance variations
%in individual clusters, that are associated to the light element variations observed also spectroscopically.

By taking advantage of both spectroscopy and photometry, it has been definitively established that individual GCs
host roughly coeval multiple populations (MPs) of stars, born with chemical abundance distributions
characterised by anticorrelations between C-N and O-Na (sometimes also Mg-Al) pairs, and a range of He abundances.
For the majority of Galactic GCs the abundances of the other elements, and in particular Fe,
  are remarkably uniform within individual clusters 
  \citep[see, e.g., the reviews by][]{gcb:12, bl18, gratton:19},
  although it does exist a small sample of objects like $\omega$~Cen, Terzan
  5, or M2, known to host stars born with a range of initial Fe abundances\footnote{Interestingly, it has
    been found that within these \lq{peculiar}\rq\ clusters C-N, O-Na anticorrelations are present
    among stars with the same Fe abundance \citep[see, e.g.][]{marino:11}.}.

Most scenarios for the origin of MPs
\citep[reviewed, e.g., in][]{bl18} invoke subsequent episodes of star formation.
Stars with CNONa (and He) abundance ratios similar
to those observed in the halo field populations are supposed to be the first objects to form (we denote them as P1 stars),
while stars enriched in N and Na (and He) and depleted in C and O formed later (we denote them as P2 stars), 
from freshly synthesised gas ejected by some class of polluter stars belonging to the P1 population.
To date, none of the proposed scenarios is able to explain quantitatively the ensemble of chemical
patterns observed in individual GCs \citep[see][for independent discussions]{r15, bl18}. Also, recent indications  
of He abundance variations among P1 stars in a sample of GCs \citep[][]{lsb18, He18}
is particularly difficult to accommodate by these scenarios.

Photometric \citep[see, e.g.,][and references therein]{larsen14, dale16, gilligan19, lagioiab, nardiello19}
and to a lesser extent spectroscopic \citep{m09} observations  
have also shown that this MP phenomenon --meaning that individual clusters host stars 
born with a range of chemical abundances displaying the same anticorrelation patterns as
in Galactic GCs-- is not confined to Galactic GCs, for 
MPs have been discovered also in  
old clusters of the Magellanic Clouds, the Fornax dwarf galaxy, and M31.
Integrated spectroscopy of clusters in  M31 also reveal the signature of MPs 
amongst old massive clusters in this galaxy \citep[see, e.g.,][]{s13}.

Additionally, recent spectroscopic and to a much larger extent photometric studies of
small samples of intermediate-age, 
resolved extragalactic massive clusters, have shown that  clusters 
down to ages of $\sim$2~Gyr do host MPs \citep[see, e.g.,][and references therein]{h19,lagioia19,mart19}.
This result adds an additional and very important piece of information to the MP puzzle, because it
strongly suggests 
a close connection between the formation of old GCs and young massive clusters \citep[see, e.g.,][]{kr}. 

It should be clear from this very brief summary, that photometry has played and continues to play
a crucial role in the detection and
characterization of MPs
in resolved clusters. Obviously, high-resolution spectroscopy enables a more detailed
investigation of the chemical patterns of P2 stars, and has
been responsible for the discovery of the MP phenomenon, 
but for the majority of Galactic GCs this type of 
analysis is restricted to the bright red giant branch (RGB) stars, and in any case to a
limited number of objects. Even harder is to perform spectroscopy of individual stars in extragalactic objects.

On the other hand, imaging allows us to disentangle efficiently P1 and
P2 populations in individual clusters, through photometry of samples of many thousands of stars, 
and to cover wider regions in the sky compared to spectroscopic surveys.
This has enabled us to infer robustly the number ratios of P1 to P2 stars, and study their radial distributions,
by virtue of the large samples of objects observed. 
For example, the HST {\sl UV Legacy Survey of Galactic Globular Clusters} \citep{p15} has taken advantage
of HST photometric high precision and accuracy, to detect and characterize 
MPs in a large sample of Galactic GCs, and explore the link
with their host cluster properties with unprecedented precision. HST photometry has also been pivotal in
the discovery of  MPs in several intermediate-age massive clusters \citep[see, e.g.,][]{mart19}.
The information gained from photometry is therefore
crucial to try and identify the mechanisms for the formation of massive star clusters.

The purpose of this review is to discuss the theoretical underpinning and application of the main photometric
tools to able to identify and characterize MPs in resolved star clusters.
We first summarize in Sects.\ref{sec:scenario} and \ref{SED} 
the impact of P2
chemical abundances on theoretical stellar models and isochrones, and the predicted spectral energy
distributions.
These results are then employed in Sects.~\ref{chrom}, \ref{He}, and \ref{Heb} to introduce and discuss 
several photometric diagnostics to detect and characterize MPs in star clusters.
We close in Sect.~\ref{concl} with a summary of the main results related to MP discoveries and
characterization, obtained by
photometric surveys of massive star clusters, and future prospects for this type of investigation.

\section{Impact of chemical abundance anticorrelations on stellar evolution models, tracks and isochrones}
\label{sec:scenario}

\begin{figure}
\centering
\includegraphics[width=8.0cm]{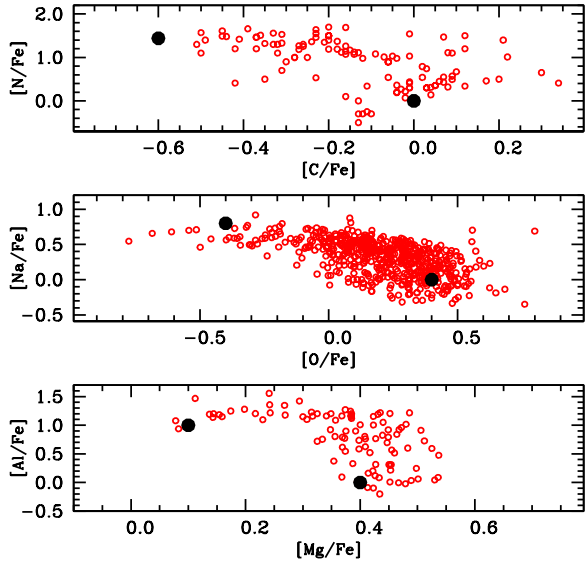}
\vskip -0.3cm
\caption{Observed  [C/Fe]-[N/Fe] ({\sl upper panel}), [O/Fe]-[Na/Fe] ({\sl middle panel}) and  [Mg/Fe]-[Al/Fe] ({\sl lower panel}) abundance patterns for stars belonging to a sample of RGB stars in Galactic GC. The filled circles display the representative P1 and P2 abundance ratios (with the same CNO sum)
  employed in our analysis (see text for details).}
\label{anticorr}
\end{figure}

As mentioned in the Introduction, in the context of massive cluster MPs are 
groups of stars born with different initial chemical compositions, but very similar ages.
The chemical pattern that tells apart the various populations comprises anticorrelations between
the abundances of specific pairs of elements amongst stars in the same cluster, namely 
C-N, O-Na, and sometimes Al-Mg (and rarely Si-F) \citep[see the discussion in][and references therein]{gratton:19}. P1 stars 
show abundance ratios typical of field stars -- with the same [Fe/H] -- in the cluster environment
(for example, field Galactic halo stars in case of Galactic GCs), while P2 stars comprise objects
with enhanced N and
depleted C, enhanced Na and depleted O, enhanced Al and depleted Mg, compared to the P1 population.
Clusters with spectroscopic measurements of C, N and O, show that the sum ${\rm (C+N+O)}$ 
is constant within current uncertainties (a factor 1.5-2), with few exceptions represented by the
Galactic GCs NGC~1851 \citep[but see][for conflicting results on this cluster]{villanova:10, yong:09},
NGC~6656 \citep{marino:12}, and $\omega$~Cen \citep{marino:12c}. This latter cluster is also well known to display
a large range of [Fe/H].
%The evidence that the sum  ${\rm C+N}$ increases as well with decreasing C, is a proof that the {\sl C-N anti-correlation} is not simply due to the conversion of C into N as a %consequence of the {\sl CN cycle} being at work. 
Figure~\ref{anticorr} shows abundances of RGB stars for a small sample of Galactic GCs, which  
follow clear C-N, O-Na and Mg-Al anticorrelations 
\cite[see][]{cmpsf:13}.

%Both spectroscopic and photometric empirical findings clearly support the conclusion that these peculiar chemical abundance patterns have to be coupled with - from moderate to %very high - helium enhancements in SG stars with respect the FG.

The pairs of elements involved and their anticorrelated patterns points to an origin linked to high-temperature proton captures
during CNO H-burning. This, in turn, suggests also the presence of He abundance variations between P1 and P2 stars. 
Helium variations ($\Delta Y$, where $Y$ denotes the He mass fraction) can be determined by direct spectroscopic
measurements in bright RGB stars (through a chromospheric line) and horizontal branch (HB) stars hot enough to show He
photospheric lines ($T_{eff} >$8500~K), but cooler than the $T_{eff}$ limit for the onset 
of atomic diffusion \citep[$\sim$12000~K, see e.g.][]{da13, marino14, mucc14, p11, villanova12}. The few available
measurements 
are consistent with some He-abundance spreads coupled to the CNONaMgAl anticorrelations, although the uncertainties are large.

To employ photometry as an additional diagnostic to detect and characterize 
cluster MPs, we need to assess the impact
of the P2 chemical abundance patterns first on stellar evolution models, and the corresponding tracks and
isochrones in the Hertzsprung-Russell diagram (HRD), and then on the predicted spectral energy distributions (SEDs) .

\subsection{Impact of the light-element anticorrelations on stellar models and evolutionary tracks}
\label{light}

The effect of CNONaMgAl anticorrelations on stellar models has been investigated by \cite{salaris:06}, \cite{pietrinferni:09},
\cite{cmpsf:13}, and \cite{dotter}.
Here we first explore the effect of this metal abundance patterns at fixed initial helium abundance.

Reference calculations for the P1 populations are from the BaSTI \citep{pietrinferni:06} models computed with an
$\alpha$-enhanced [$\alpha$/Fe]=0.4 metal mixture.  
The P2 models --unless otherwise noted-- 
%with just the CNONa anticorrelations
have been calculated, at a given [Fe/H] and $Y$, including depletions of C, O and Mg by 0.6~dex
0.8~dex and 0.3~dex by mass, respectively, and enhancements of N, Na and Al by  1.44~dex, 0.8~dex and 1~dex,   
compared to P1 abundances \cite{cmpsf:13}.
%\citep{sswc:11}.
%To these changes, \cite{cmpsf:13} also added an enhancement of Al  by 1~dex, and a depletion of Mg by 0.3~dex, to explore the impact of
%the Mg-Al anti-correlation, that however is not detected in all clusters.
%In the following we refer to this P2 chemical patterns as the
%CNONaMgAl mixture.
The (C+N+O) sum in the P2 metal composition is within 0.5\% of the P1 value, and the resulting total metal mass fraction $Z$ is
the same between P1 and P2 models at a given [Fe/H] and $Y$.
These P2 abundance patterns are shown in Fig.~\ref{anticorr}, superimposed on the abundances observed
in a sample of Galactic GCs. 
We chose in our modeling somewhat {\sl extreme} values of the observed anticorrelations 
\citep[as observed in NGC~6752 and NGC~2808, see also the detailed discussions in][and references therein]{carretta:05,gcb:12,gratton:19}
to maximise their impact on  
stellar models as well as on their SEDs (see Sect.~\ref{SED}).

%The adopted chemical abundances are shown in fig.~\ref{anticorr}, and the one selected as representative of the SG chemical pattern is similar to the most extreme cases %observed in clusters like NGC~6752 and NGC~2808. As already mentioned, this choice has been done with the aim to maximize the impact of the SG composition on both the stellar %models and spectral properties.
%The selected SG abundance ratios correspond to [(C+N)/Fe]=0.73 (compared to [(C+N)/Fe]=0.0 for the 
%FG $\alpha$-enhanced mixture), [(C+N+O)/Fe]=0.37 (within 0.5\% of the FG value), 
%[(Mg+Al)/Fe]=0.28 (compared to [(Mg+Al)/Fe]=0.38 for the FG mixture).

In addition, we also consider the effect of a CNO-enhanced (${\rm CNO_{enh}}$) metal distribution, with the (C+N+O) sum larger by a factor of 2 
compared to P1 abundances \citep[the only difference compared to the CNO-constant P2 composition is
  that N is enhanced by 1.8~dex rather than 1.44~dex, see][]{sswc:11}.
This is to investigate the effect of a 
CNO-enhancement, as observed in some Galactic GCs, like NGC~6656 and $\omega$~Cen.
In this case the metallicity $Z$ of P2 models is a factor of 2 higher than the P1 counterpart with the same [Fe/H] and $Y$.

From basic stellar physics we expect the differences between P1 and P2 composition to affect the radiative Rosseland mean opacity $\kappa$, the equation of state and the energy generation efficiency, which in turn can have an impact on luminosity, $T_{\mathrm eff}$ and lifetime of the models.
Actually, the reference P2 composition with unchanged CNO sum does not have any significant impact on model structure, lifetime, evolutionary tracks, in the regime of Galactic GCs 
\citep[see, e.g.,][]{cmpsf:13}, but also, as we have verified with appropriate calculations, at intermediate ages down to 1-2~Gyr and [Fe/H] up to about half solar values.
 
\begin{figure}
\centering
\includegraphics[width=6.5cm]{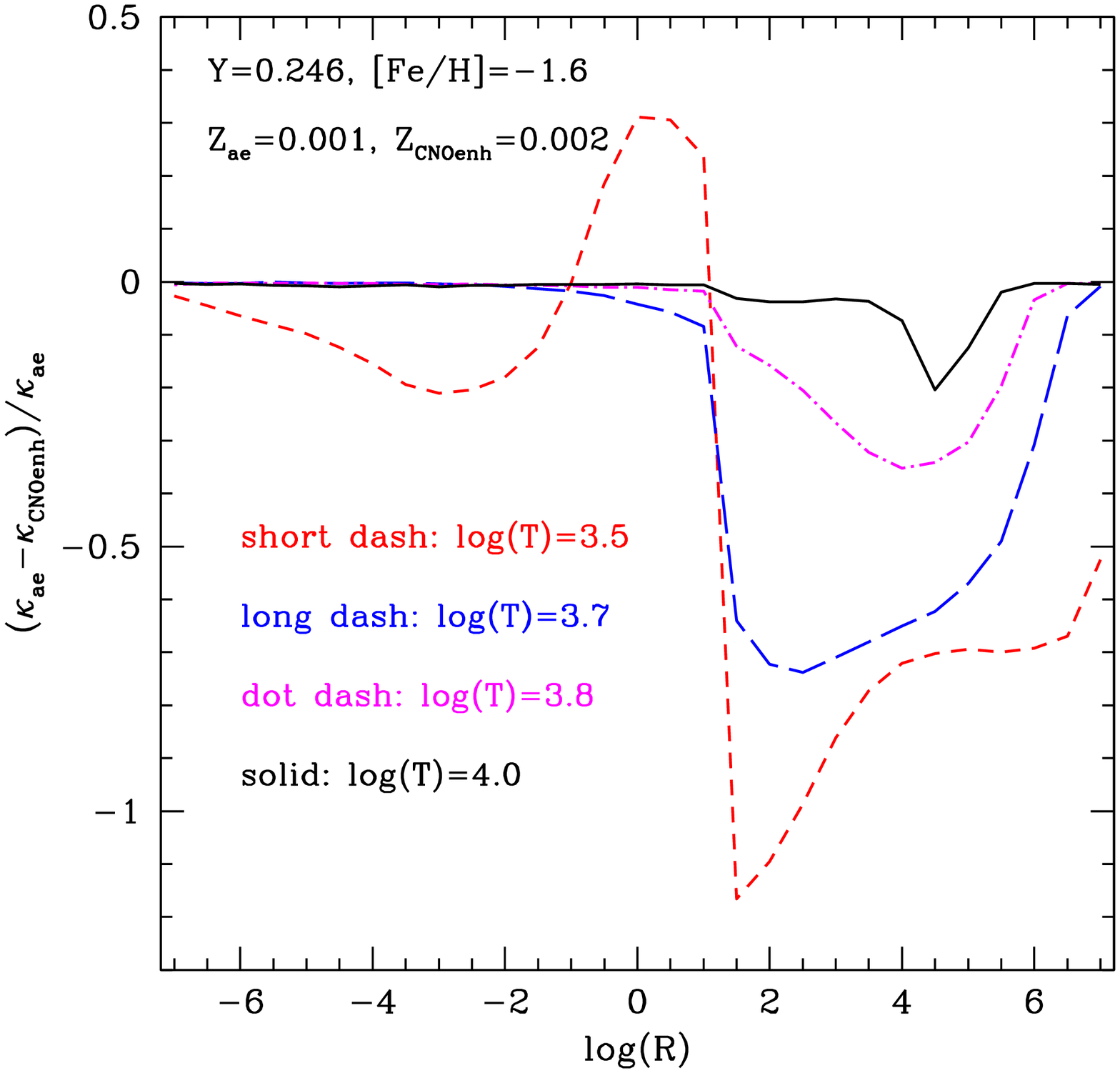}
\includegraphics[width=6.5cm]{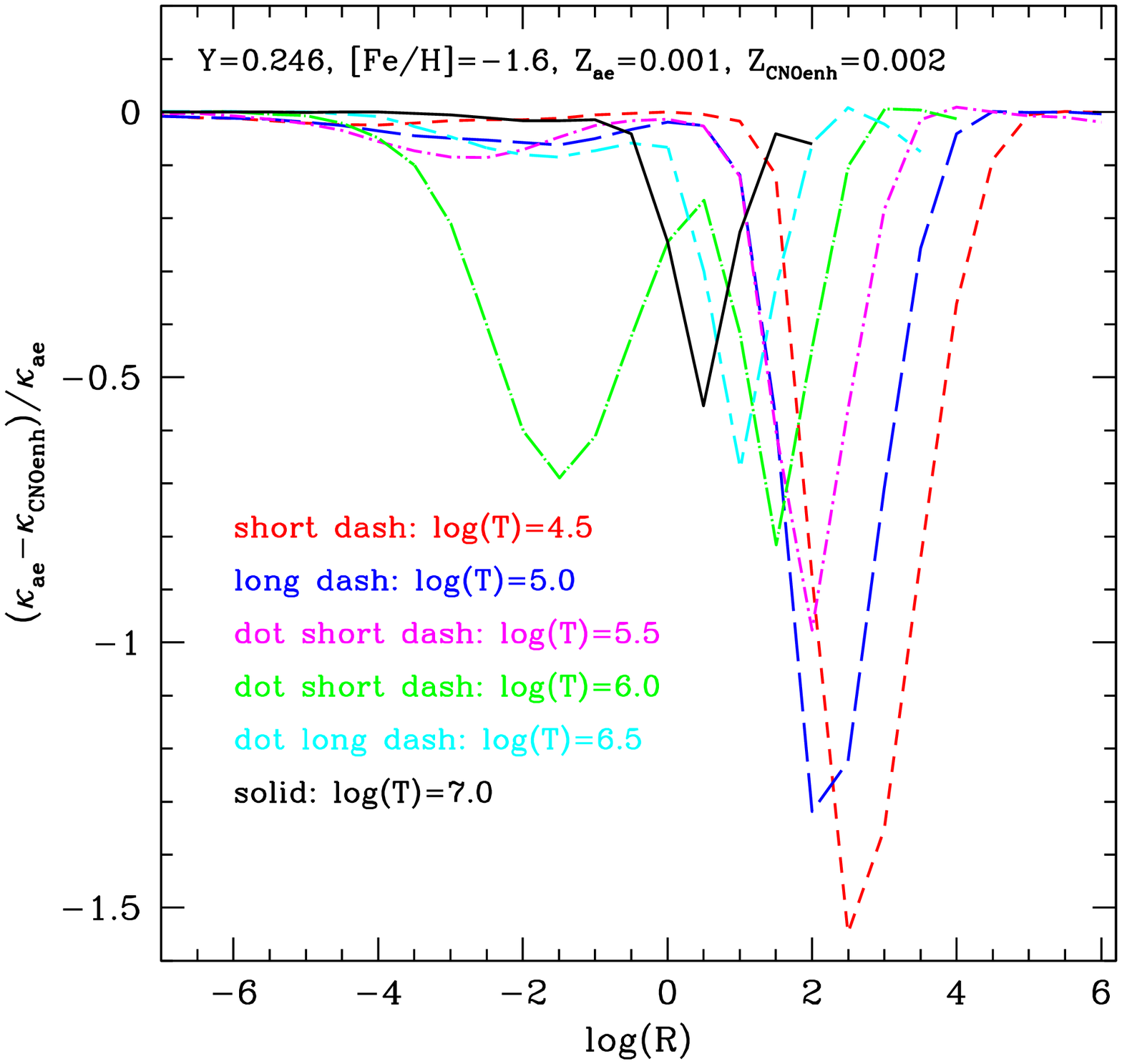}
\vskip -0.3cm
\caption{Relative difference of the radiative Rosseland mean opacity 
between a standard P1 $\alpha-$enhanced composition, and the CNO-enhanced P2 ${\rm CNO_{enh}}$  
composition at fixed [Fe/H] and $Y$, and several temperatures (see labels), as a function of the density parameter R
(see text for details).
The corresponding total metal mass fractions $Z$ are also labelled.}
\label{figopacno}
\end{figure}

A sizable effect is however expected in case of the ${\rm CNO_{enh}}$ P2 composition \citep{salaris:06, cassisi:08, pietrinferni:09, ventura:09, don:12}, mainly through opacity and energy generation efficiency.
Figure~\ref{figopacno} displays the relative difference of the radiative opacity in both high- and low-temperature regimes, between matter with a normal $\alpha-$enhanced P1 heavy element distribution and matter with the P2 ${\rm CNO_{enh}}$ mixture, for [Fe/H]=$-$1.6 and $Y$=0.246. The differences are displayed as a function of the density parameter R=$\rho$ /$T_{6}^{3}$
--where $\rho$ is the density in ${\rm g\cdot{cm^{-3}}}$ and $T_6$ the temperature in million Kelvin-- and various temperatures $T$

\begin{figure}[t]
\centering
%\subfigure{
\includegraphics[width=6.5cm]{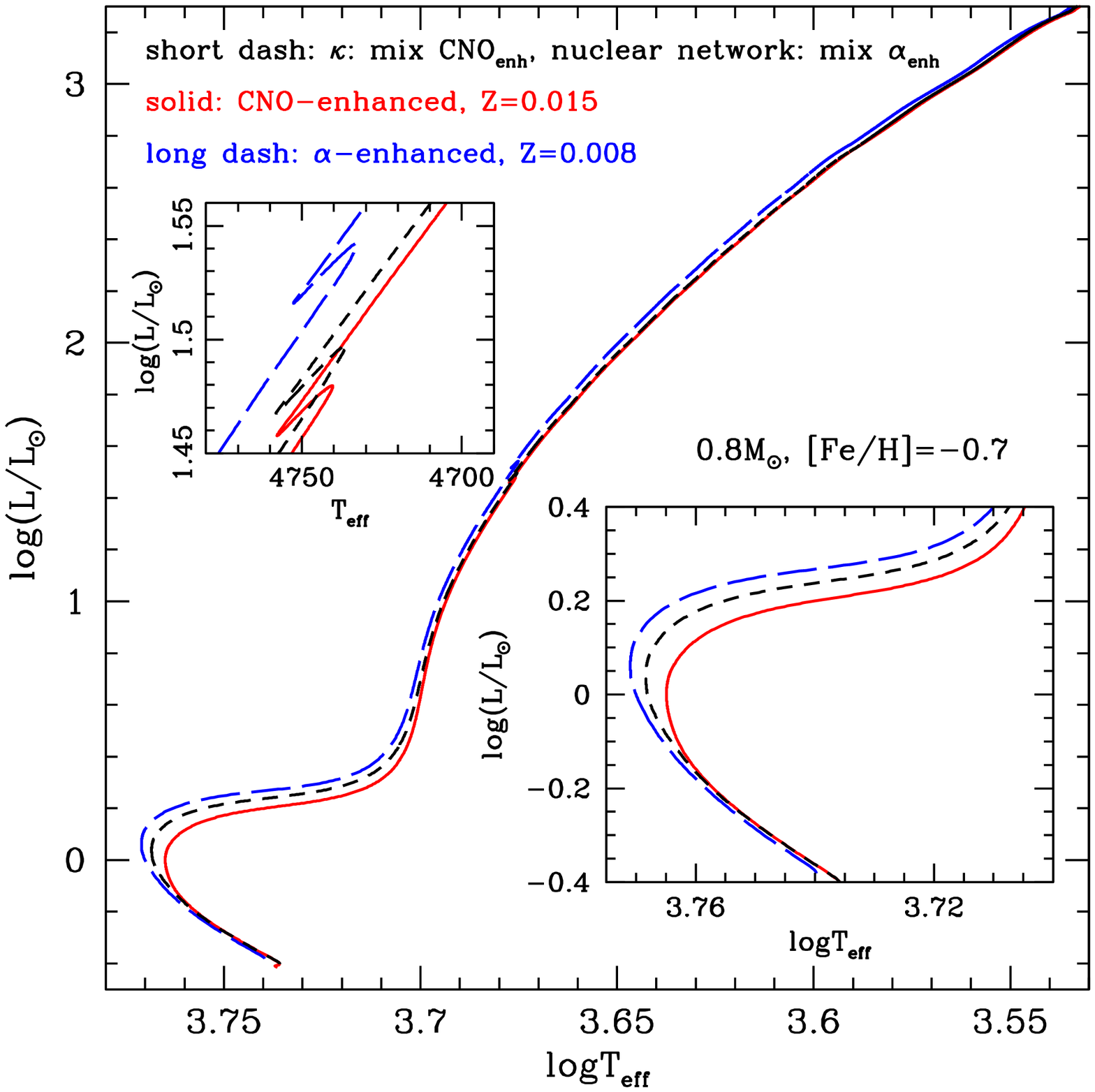}
%}
%\subfigure{
\includegraphics[width=6.5cm]{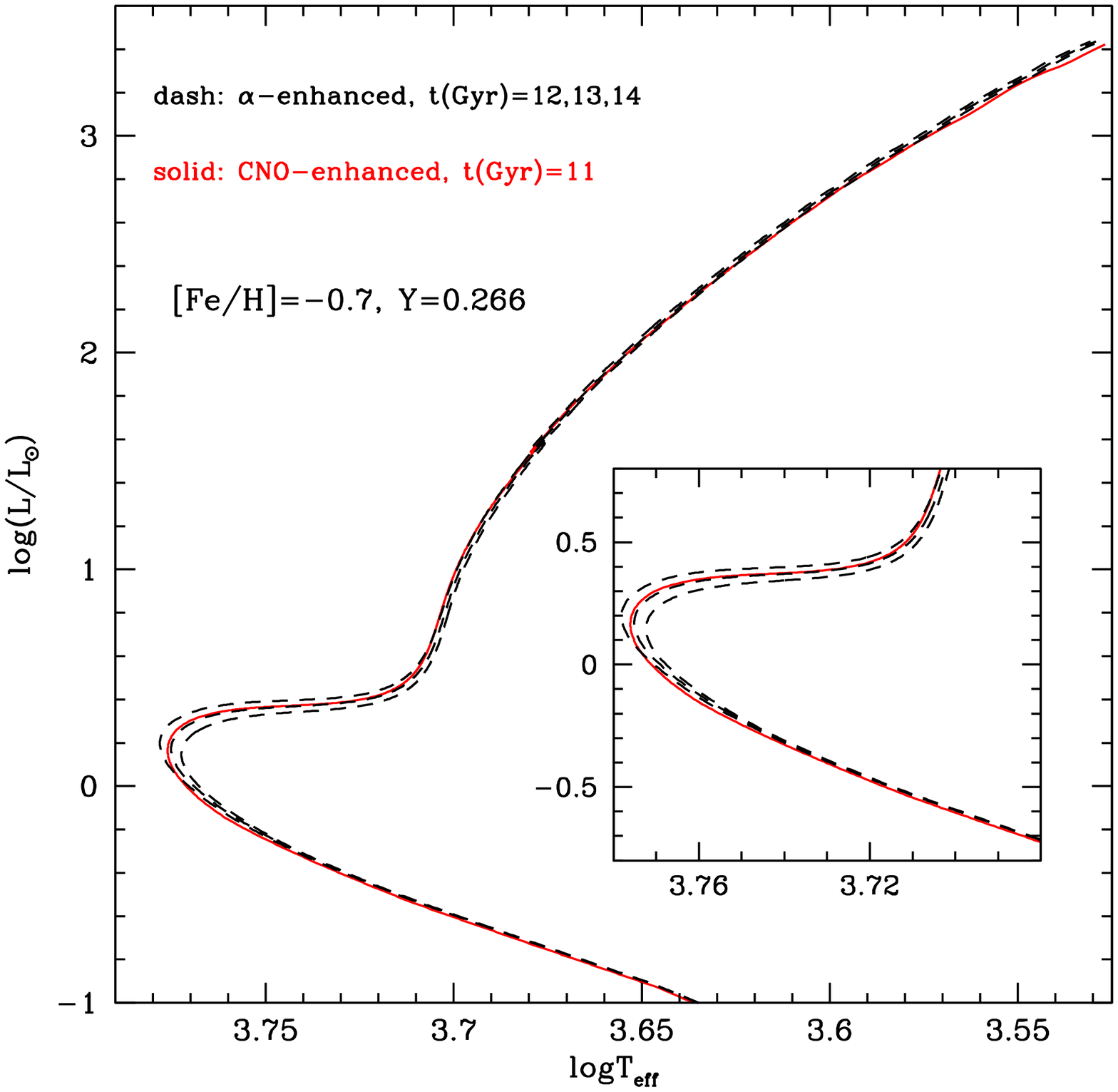}
%}
\vskip -0.3cm
\caption{{\sl Upper panel}: HRD of evolutionary tracks for a $0.8M_\odot$ model computed with the $\alpha$-enhanced P1 composition, and the P2 ${\rm CNO_{enh}}$ metal mixture, 
both with the same [Fe/H] ([Fe/H]=$-$0.7) and initial He abundance ($Y$=0.266).
The short-dashed line shows a track calculated with radiative opacities appropriate for the P2 composition, but the same CNO abundances of the P1 composition 
in the nuclear burning network (see text for details).
% a numerical experiment performed by using the radiative opacity appropriate for the ${\rm CNO_{enh}}$ mixture and the nuclear network appropriate for the $\alpha-$enhanced, %reference mixture. 
%The aim of this experiment is to disentangle the contribution of radiative opacity and of the nuclear burning efficiency
%in determining the morphology  in the H-R diagram of the evolutionary tracks for SG stars with an enhanced CNO sum. 
The two insets show the RGB bump and the TO-SGB regions, respectively. 
{\sl Lower panel}: HRD of theoretical isochrones computed with the same P1 and P2 compositions of the upper panel,
and various ages. The inset shows the TO-SGB regions.}
\label{figtrkcno}
\end{figure}

For temperatures lower than about 10,000~K and ${\rm \log{R}\le -1.0}$, i.e. in the regime typical of the envelopes of low-mass models along the
main sequence (MS), subgiant branch (SGB) and RGB, differences are small, except for very low temperatures around 3000~K.
On the other hand, at temperatures above 10,000~K and larger values of $R$, typical of model interiors, differences are much more significant.
The opacity of the ${\rm CNO_{enh}}$ composition is typically larger than that of the P1 metal mixture, with differences mainly confined to R values 
in the range between ${\rm \log{R}\sim +1}$ and $\sim +4$. 
The larger values of the opacities in this regime are consistent with the fact that C, N and O are among
the main contributors  
to the high-temperature opacity of the stellar matter \citep[see, e.g.,][]{scs93, don:12}. 

\begin{figure}
\centering
\includegraphics[width=7.0cm]{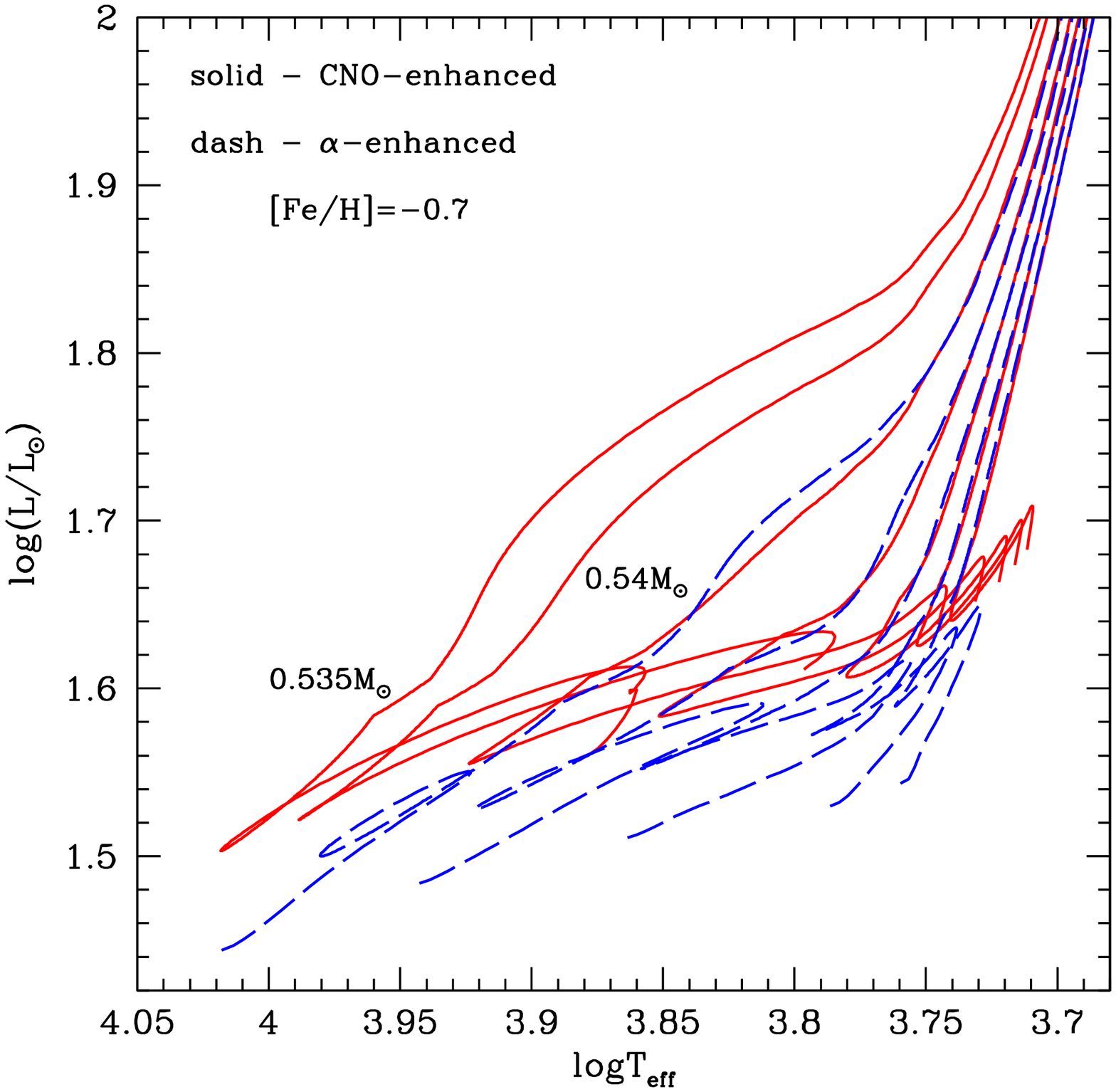}
\vskip -0.3cm
\caption{HRD of several HB evolutionary tracks with the same P1 and P2 compositions of Fig.~\ref{figtrkcno}, and varying mass. 
The values of the lowest masses for the two sets of tracks are labelled. The 
other tracks correspond to masses (increasing towards lower effective temperatures) ranging from 0.55 to 0.59$M_{\odot}$, in steps of 0.01$M_{\odot}$.}
\label{fighb}
\end{figure}

Results from the computation of low-mass models with P1 and P2 compositions can be summarised as follows:
 
\begin{itemize}

\item If the CNO sum is the same in P1 and P2 models, evolutionary tracks (including the core He-burning stage) and isochrones of P2 populations are identical to their
  $\alpha$-enhanced P1 counterparts with the same [Fe/H] and $Y$, meaning that \emph{there is no need for dedicated calculations}.
This is important, given that, as already mentioned, measurements (albeit not very extensive) of the CNO sum point to a constant value in P1 and P2 stars, within the measurement errors.
%This is very convenient, relevant advantage when considering the impressive variety of light element chemical patterns observed in the context of the multiple population %phenomenon;

\item if the CNO sum is enhanced in P2 models, the morphology of evolutionary tracks in the HRD is modified, as shown in the upper panel of Fig.~\ref{figtrkcno} for the MS, SGB and RGB.  
For a given initial mass (at fixed [Fe/H] and $Y$), both MS turn off (TO) and SGB are fainter (the TO is also cooler) in comparison with standard $\alpha-$enhanced P1 tracks. 
This behaviour is due to both the increased efficiency of the CNO-cycle in CNO-enhanced models, and the increase of $\kappa$.
The variation of the TO luminosity and $T_{\mathrm eff}$ is mainly due to the increased efficiency of the CNO cycle, as shown by the results of the numerical experiment in Fig.~\ref{figtrkcno}. If the CNO enhancement is accounted for only in the opacities, the effect on the TO position in the HRD is more than halved.  

\item The evolutionary lifetimes of the CNO-enhanced P2 models are affected by the increase of the CNO sum due to the altered efficiency of the CNO-cycle.
  However, for masses typical of stars currently evolving in Galactic GCs, due to the marginal contribution of the CNO-cycle  
  to the energy budget, this effect is small, amounting to $\sim1$\% for the 0.8$M_\odot$ tracks shown in Fig.~\ref{figtrkcno}, 
the P1 $\alpha-$enhanced models being younger by about 120~Myr.
Whilst the $T_{\mathrm eff}$ of the RGB is basically unchanged, the RGB bump luminosity becomes fainter in the P2 track 
by $\Delta\log(L/L_\odot)\sim 0.06$. Again, as for the TO, the main effect is the increased efficiency of the CNO cycle (see Fig.~\ref{figtrkcno}).

\item Despite the fact that during the first dredge-up (FDU) the convective envelope reaches deeper layers in ${\rm CNO_{enh}}$  models, the amount of helium dredged to the surface is exactly the same ($\Delta Y = 0.017$) as in P1 calculations. This is due to the increased efficiency of the CNO cycle in ${\rm CNO_{enh}}$ models, which confines core H-burning to
  regions closer to the centre.

The properties of stellar models at the He-flash are also changed in the ${\rm CNO_{enh}}$ calculations. In the case of the tracks displayed in Fig.~\ref{figtrkcno},  
the He-core mass at the tip of the RGB (TRGB) is equal to $M_{cHe} = 0.4687M_\odot$ for the P2 composition, while the corresponding P1 value is $M_{cHe} = 0.4734M_\odot$. This difference translates to a $\Delta\log(L/L_\odot)\sim 0.015$ difference of the TRGB luminosity.

\item Despite the lower value of $M_{cHe}$ and the larger total metallicity $Z$, at fixed [Fe/H] and initial $Y$ the ${\rm CNO_{enh}}$ HB tracks and
  their Zero age HB (ZAHB) location are brighter than the P1 counterparts, as shown in Fig.~\ref{fighb}.
The reason is the increased efficiency of the CNO-burning in the hydrogen shell, 
as a consequence of the increased CNO sum. The brightness difference along the ZAHB at
$\log(T_{\mathrm eff})$= 3.83 --taken as 
representative of the mean effective temperature of the RR Lyrae instability strip-- is equal to
$\Delta\log(L/L_\odot)\sim 0.05$, for models with [Fe/H]=$-$0.7. 
%This luminosity difference decreases with decreasing 
%metallicity, as can be appreciated by looking at the ZAHB loci for ${\rm [Fe/ H]=-1.6}$.
For a fixed total mass, the ZAHB location of CNO-enhanced P2 models is cooler compared to P1 models: 
However, as a consequence of the higher efficiency of the H-burning shell, 
P2 models display more extended blue loops in the HRD, as shown in Fig.~\ref{fighb}.

\item The HRD of MS-SGB-RGB isochrones with P2 ${\rm CNO_{enh}}$ composition are identical to P1 results, 
  but for the TO and SGB regions, which are fainter
  (the TO is also cooler) 
  for the P2 composition, as shown in Fig.~\ref{figtrkcno}. At the [Fe/H] and $Y$ values of this comparison,
  the 11~Gyr P2 CNO-enhanced isochrone is perfectly mimicked by a $\sim 1.5-2$~Gyr older P1 isochrone.

\end{itemize}

\subsection{Impact of helium enhancement}
\label{sub:helium}

The initial helium abundance has a major effect on the structure and evolution of stellar models,  
and this section discusses in detail the impact of
  helium abundance variations on the shape and location of isochrones in the HRD.
  Section~\ref{sub:vlm} will focus on the lower main sequence, Sect.~\ref{sub:rgb} on RGB models and
  Sect.~\ref{sub:hb} on the core He-burning phase.

\begin{figure*}
\centering
\includegraphics[width=8.0cm]{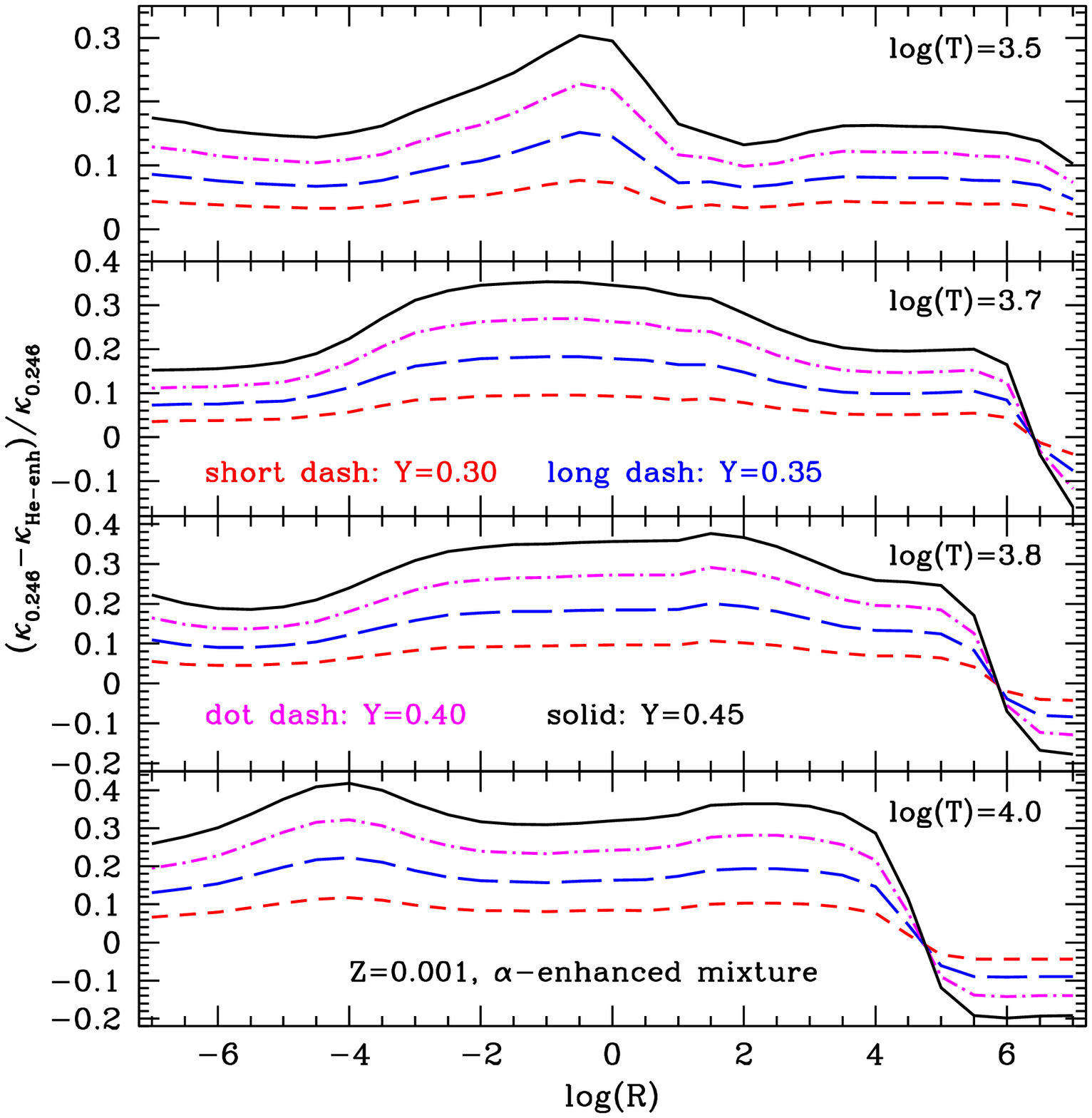}
\includegraphics[width=8.0cm]{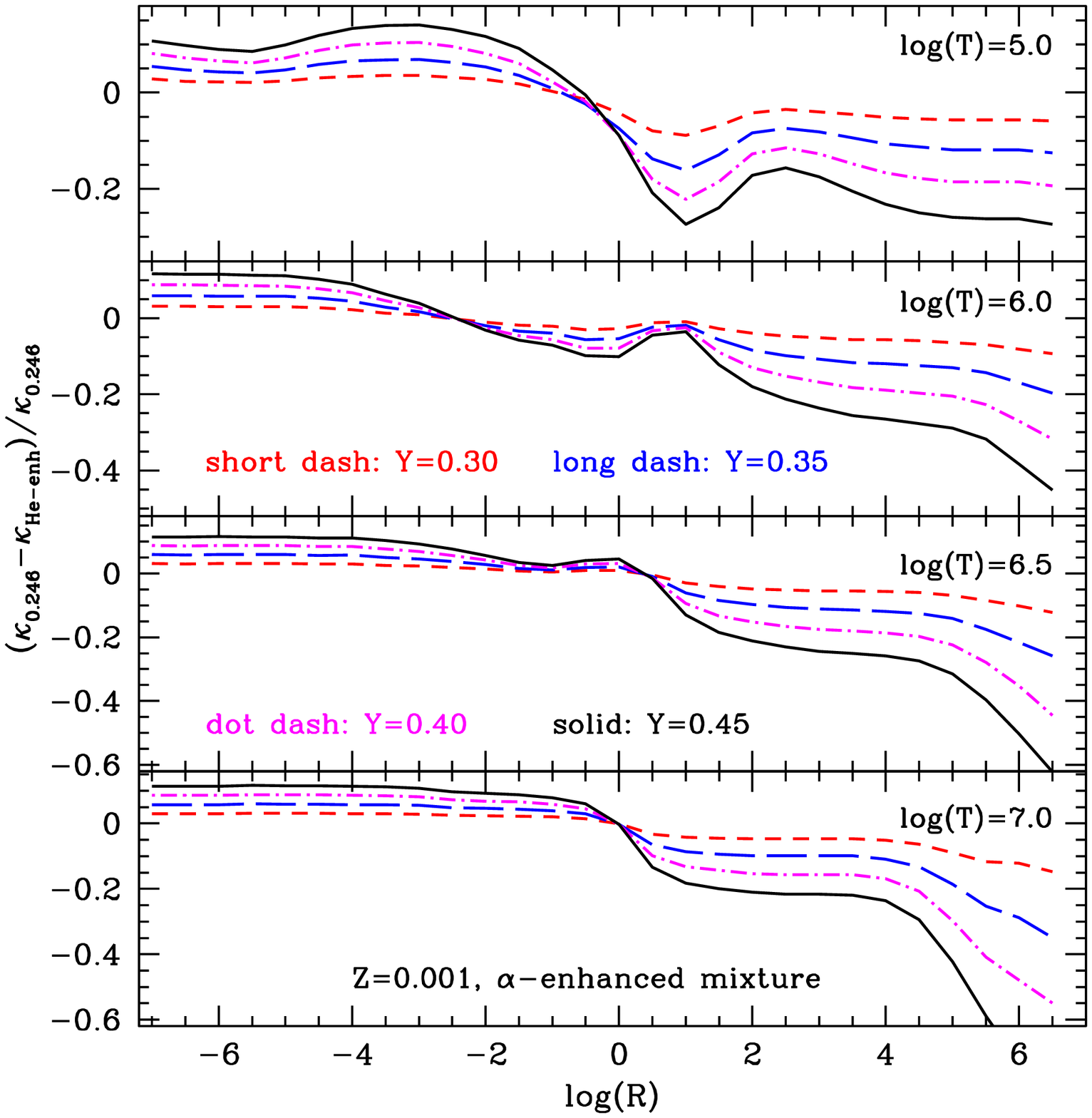}
\caption{{\sl Top panel}: Relative difference between the radiative opacity for a chemical composition with $Y$=0.248, and He-enhanced composition (see labels) at fixed metallicity ($Z$=0.001), as a function of the density parameter R for the labelled temperatures. {\sl Bottom panel}: As the top panel, but in the high-temperature regime. The opacity calculations in the low-temperature regime are from \cite{ferguson:05}, while those in the high-temperature range
come from \cite{opal}.}
\label{figopahe}
\end{figure*}

  Let us at  first discuss the impact of a He abundance change on the opacity of stellar matter.
An increase of the He mass fraction $Y$ at fixed metallicity causes a reduction 
of the radiative opacity, as shown in Fig.~\ref{figopahe}. In this figure we show the relative difference 
of the Rosseland mean radiative opacity between a a composition with standard $Y$=0.248 and compositions with various He
enhancements,
for a representative metallicity $Z$=0.001.
At temperatures below or equal to 10,000~K, an increase of $Y$ decreases 
$\kappa$ at fixed R. At $\log (T)$=3.8
the decrease is on average on the order of $\sim5$\% when going from $Y$=0.248 to $Y$=0.30, but increases  
to $\sim30$\% when $Y$=0.45. 
Only for R larger than $\sim5$ 
the trend is reversed and the opacity increases when increasing the He abundance\footnote{In the interiors of low-mass models, ${\rm \log{R}}$ reaches values on the order of $\sim0.5$
  only during the RGB advanced evolutionary phases. However in this regime the electron conduction opacity is the dominant contributor to 
  the opacity in the He-core.}.
This behaviour is confirmed also at temperatures above 10,000~K --although the change of the opacity is smaller in comparison to lower temperatures-- but the value of R beyond which the trend is reversed is shifted to lower values, as the temperature increases.
This general decrease of $\kappa$ when increasing the He content at fixed $Z$ is due to the decrease of the hydrogen mass fraction $X$, hydrogen being 
a major opacity source via the ${\rm H^-}$ ion. The decrease of $\kappa$ makes He-enhanced stellar models generally hotter than models with standard $Y$.

The opacity reduction in He-enhanced models makes them also brighter during the MS. An additional and larger increase of the luminosity 
in this phase is caused by the change of the mean molecular weight\footnote{In case of complete ionisation, the mean molecular weight is given by $\mu=\frac{1}{2X+\frac{3}{4}Y+\frac{Z}{2}}$.} $\mu$ of the stellar matter when the He abundance changes.
In fact, the H-burning luminosity ${\rm L_H}$ increaes with $\mu$ as ${\rm L_H\propto\mu^7}$. When He increases at fixed $Z$, the mean molecular weight also increases, and this translates to a larger H-burning luminosity. Given that 
${\rm \Delta{L_H}/L_H=7\Delta\mu/\mu}$, an increase $\Delta Y=0.10$ 
%--lower than the maximum He enhancements expected for the bluest MS stars in the GCs $\omega$~Cen and NGC~2808 -- 
causes a $\sim 50$\% increase of the H-burning luminosity. The combined effect of the radiative opacity decrease and the increase
of H-burning efficiency 
make He-rich stellar models brighter and hotter along the MS. As a consequence, their MS lifetime (${\rm t_H}$) is significantly reduced:
for a $0.8M_\odot$, ${\rm t_H}$ decreases from 11.3~Gyr to 3.8~Gyr when $Y$ increases from 0.246 to 0.40. 

\begin{figure*}
\centering
\includegraphics[width=8.0cm]{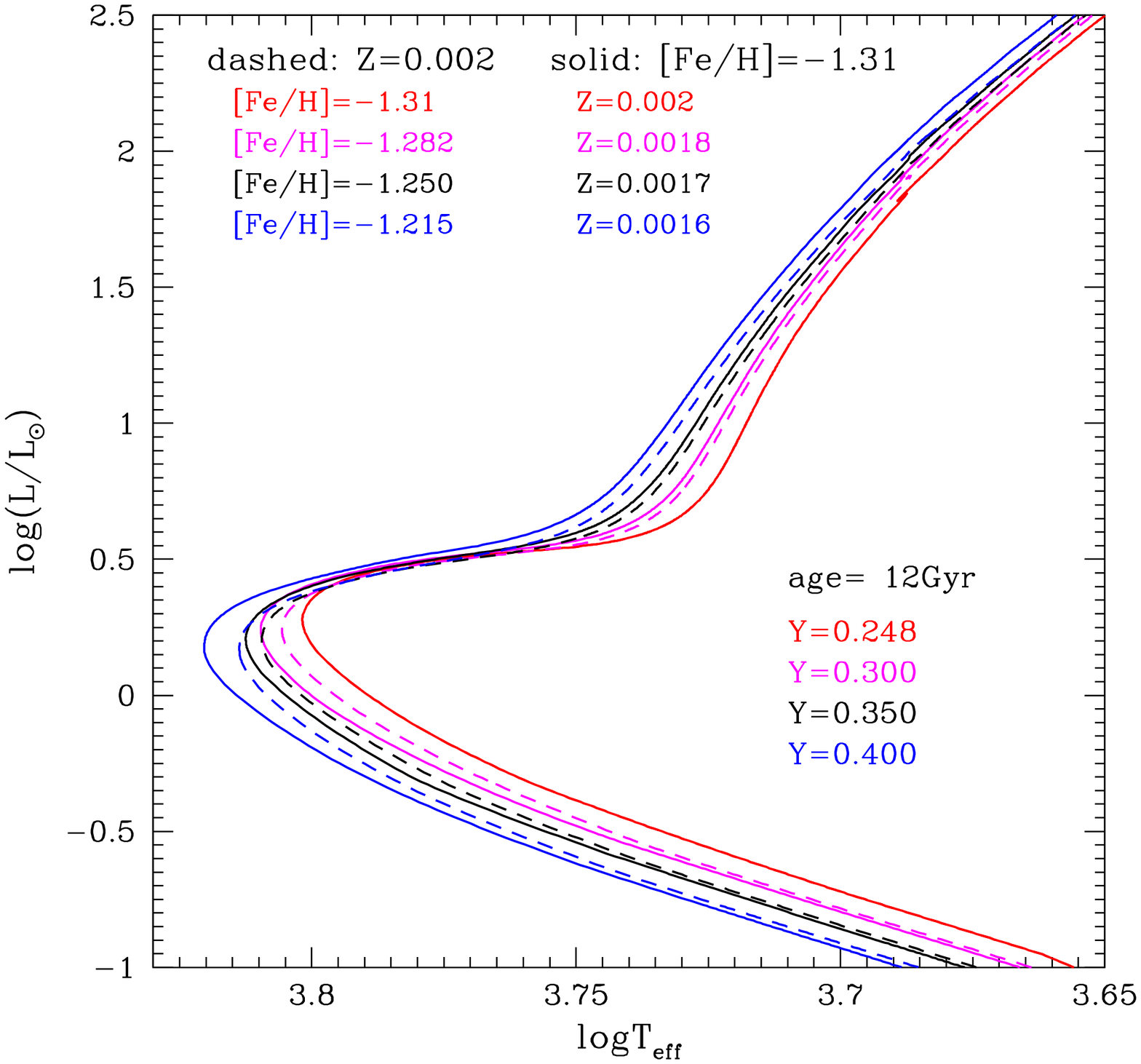}
\caption{Theoretical isochrones from the MS to the RGB, for an age of 12~Gyr 
and various assumptions about the initial He content, computed by alternatively assuming a constant metal mass fraction $Z$ 
(dashed lines) or a constant [Fe/H] (solid lines -- see text for details).
The different values of [Fe/H] corresponding to varying $Y$ at fixed $Z$ are displayed, together with the values of
$Z$ corresponding to varying $Y$ at fixed [Fe/H].
%{\sl Right panel}: as left panel but in the optical ACS@HST CMD ${\rm (M_{F814W}, F606W - F814W)}$ }
}
\label{figisohe}
\end{figure*}

Figure~\ref{figisohe} shows various isochrones from the MS 
to the RGB at fixed $Z$ and age, but different values of the initial He abundance. 
The most relevant features are the following: 

\begin{itemize}

\item  The various MS isochrones run parallel in the luminosity interval from the TO down to the lower mass limit in the figure.
  At fixed age, the TO becomes fainter when increasing the He content; this is a consequence of the shorter MS 
  lifetimes of He-rich stars, which imply 
  a lower mass at the TO ($M_{\mathrm TO}$). For a 12~Gyr old isochrone $M_{\mathrm TO}$ is equal to $0.806M_\odot$ when $Y$=0.245,
  and $0.610 M_\odot$ for $Y$=0.40.
  This has important implications for the morphology of the HB of He-rich stellar populations;  

\item The $T_{\mathrm eff}$ of the MS at fixed luminosity increases with increasing initial He abundance,
$\frac{\Delta T_{\mathrm eff}}{\Delta Y}\sim 2.3\times 10^3 $~K; 

\item The SGB morphology is practically unaffected  by a He abundance change;

\item The $T_{\mathrm eff}$ of the RGB is also affected by a He increase: at fixed bolometric luminosity,
  the larger the initial He content,
  the hotter the RGB. The effect is larger at the base of the RGB, decreasing when increasing the luminosity.
  For example, at $\log (L/L_{\odot})$=1,  $\frac{\Delta T_{\mathrm eff}}{\Delta Y}\sim 9\times 10^2 $~K (smaller than for the MS),
  which decreases to $\frac{\Delta T_{\mathrm eff}}{\Delta Y}\sim 5\times 10^2 $~K when $\log (L/L_{\odot})$=2.

\end{itemize}

The previous analysis has been performed by comparing isochrones for various initial He abundances but
the same metallicity $Z$. Once 
the heavy element distribution is fixed, the amount of iron is set by $Z$, hence the assumption of constant $Z$ when changing the 
He abundance has the implication that [Fe/H] is not constant, because the hydrogen abundance does change. Given that 
$X+Y+Z=1$ by definition, if 
$Z$ is kept fixed, increasing $Y$ implies a decrease of $X$, hence an increase of [Fe/H].

The choice of models with different initial He abundance and constant $Z$ seems appropriate 
for MP studies, because the abundance patterns commonly observed do not 
show evidence of a significant change in the total metallicity $Z$ among the various populations in individual clusters. 
In fact, the sum of the CNO 
abundances plus the abundances of all elements not affected by the anti-correlations (the most important ones being Fe, Ne, and Si)
amount to about 90\% or more of the total $Z$. Given that the CNO sum and iron abundance are remarkably constant amongst the cluster
sub-populations --but for very few exceptions like for example $\omega$~Cen \citep{jp:10}, or 
Terzan~5 \citep{origlia:13}-- 
%and M22 \citep{marino:12}
it seems appropriate to consider He-enhancements at constant $Z$. 
In addition, when the He abundance variation amongst the cluster populations is large enough --as indeed is the case of NGC~2808-- 
it is possible to detect spectroscopically small variations of [Fe/H], consistent with the expected change
due to the H abundance decrease \citep{bragaglia:10}. 
In general, the change of [Fe/H] due to \lq{reasonable\rq} variations of $Y$ is small,
becoming more significant (${\rm \Delta{[Fe/H]\sim0.06}}$~dex or more)  only for a He increase larger than about $\Delta Y\ge$0.1. 
%we address the attention of the reader to the corresponding numbers shown in fig.~\ref{figisohe}. 

The impact on theoretical isochrones of changing $Y$ at constant [Fe/H] is also shown in Fig.~\ref{figisohe}, for
heuristic purposes.
The important point to consider is that 
to have the same [Fe/H] with an increased $Y$, the metallicity $Z$ needs to be decreased. This causes 
some important differences with respect to the case of He variations at constant metallicity. First of all, 
for a fixed He enhancement, the shift in effective temperature along the MS and the RGB
is larger for the case of constant [Fe/H], as a consequence of the lower $Z$ --hence the lower opacity $\kappa$-- that pushes
stellar models towards even hotter $T_{\mathrm eff}$ values. Also,  
the SGBs of He-enhanced isochrones at fixed [Fe/H] do not overlap well with the standard $Y$ isochrones,
as in case of constant $Z$; 
this is mainly due to the reduction of the CNO sum as a consequence of the decrease of $Z$,
which affects the CNO-cycle efficiency.

\subsubsection{He enhancement and the lower main sequence}
\label{sub:vlm}

Accurate, near-infrared photometry with the WFC3 camera on board Hubble Space Telescope
({\sl HST}) has extended the study of multiple stellar populations
in Galactic GCs to very low-mass (VLM) stars (stars with mass below $\sim$0.45 $M_{\odot}$), which populate 
the faintest portion of the MS \citep[see, e.g.,][]{milone:12c, milone:17}.
These stars are important not only to investigate the existence of differences in the mass function 
of the various cluster populations \citep[see the discussion in][]{milone:12d},
but also because VLM stars are fully convective and 
pose stringent constraint on accretion scenarios proposed to explain the MP origin 
\citep[see the discussion in][and references therein]{bastian:13,cs:14,sc:14}.

\begin{figure}
\centering
\includegraphics[width=9.0cm]{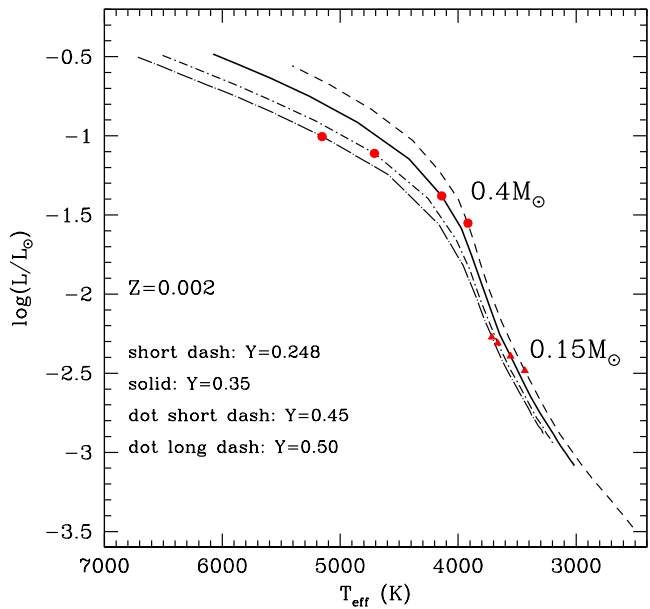}
\vskip -0.5cm
\caption{HRD diagram of 10~Gyr VLM isochrones computed at fixed $Z$ and 
the labelled initial He abundances. The points mark the location of $0.4M_\odot$ (circles), and $0.15M_\odot$ (triangles) models.}
\label{figvlm}
\end{figure}

As shown in Fig.~\ref{figvlm}, an increase of the initial $Y$ in VLM model calculations has a much smaller
effect than in more massive models.  At an age of 10~Gyr an increase $\Delta$Y=0.25 causes a change of $T_{\mathrm eff}$ by  
only $\sim250$~K, and $\Delta \log{\rm (L/L_\odot)} \sim$0.2 for a 0.15$M_\odot$ model, whereas the changes are equal to 
$\sim1200$~K and $\Delta\log{\rm (L/L_\odot)}\sim0.55$ for a $0.4M_\odot$ model.
This reduced sensitivity can be explained by the fact that VLM models are fully convective on the MS, and 
the He produced is homogeneously mixed 
throughout the whole structure, minimizing differences among models calculated with different initial He abundances.

%The impact of He enhancement on the evolution of VLM stars has the important implication that the MS splitting as due to
%the presence of distinct sub-populations with different He abundances
%should disappear - or at least significantly reduce - in optical CMDs when moving towards the fainter portion of the MS, as
%actually observed in $\omega$ Cen \cite{king:12}. 
%The MS of VLM stars is however affected in near-infrared filters by the effect of the light element (anti-)correlations on the
%bolometric corrections as it will be discussed in sect.~\ref{bolo}.

\begin{figure}
\centering
\includegraphics[width=9.0cm]{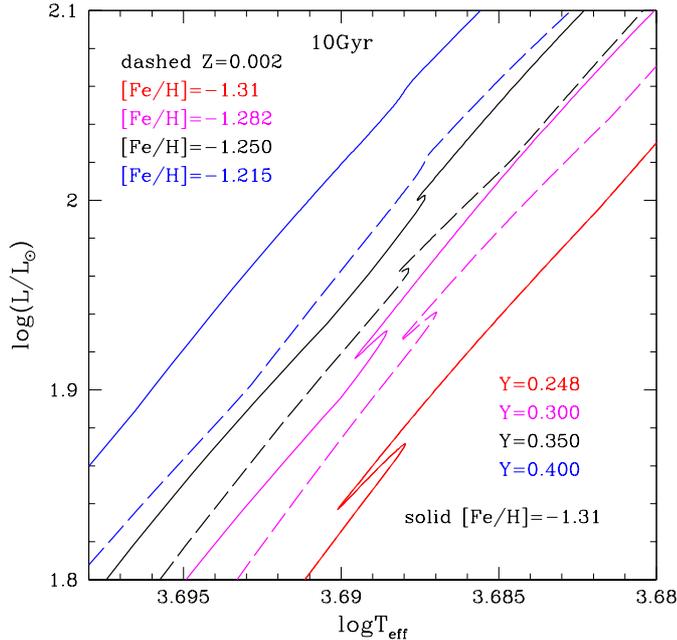}
\vskip -0.5cm
\caption{HRD around the RGB bump location for isochrones with the same labelled age, and various values of the initial He abundance.
  The isochrones have been computed under both the assumption of constant $Z$ (dashed lines) and of constant [Fe/H] (solid lines).
  The [Fe/H] values corresponding to the chosen fixed $Z$ and various $Y$ are also labelled.}
\label{figbump}
\end{figure}

\begin{figure}
\centering
\includegraphics[width=9.0cm]{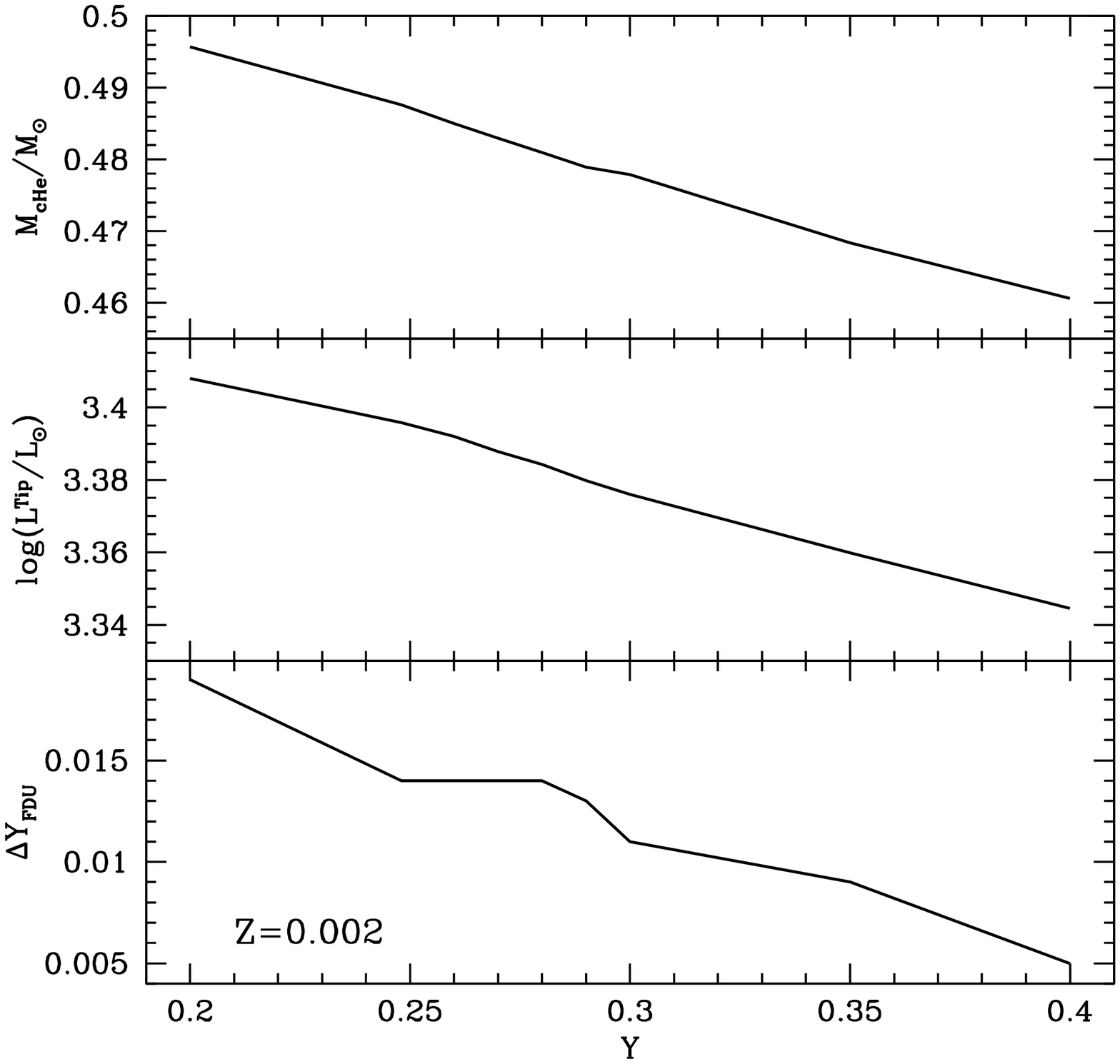}
\caption{{\sl Top panel}: The variation of the He-core mass at the TRGB for models with an age of 12~Gyr at the 
  TRGB and $Z$=0.002, as a function of the initial He abundance. {\sl Middle panel}: As the top panel, but
  for the TRGB bolometric luminosity. {\sl Bottom panel}: As the top panel, but for the increase of the surface $Y$ 
  after the FDU.}
\label{figtip}
\end{figure}

\subsubsection{Red giant branch models and He enhancement}
\label{sub:rgb}

Concerning the RGB evolution, there are three important properties affected by a change of the initial He abundance:
the $T_{\mathrm eff}$ scale, the bump luminosity, and the TRGB luminosity.

The variation of the RGB $T_{\mathrm eff}$ has been already noted
when discussing Fig.~\ref{figisohe}, and is due to the decreased opacity
when $Y$ is increased. Regarding the bump, 
we recall that the RGB bump is a local maximum in the differential luminosity function of the RGB in old populations.
It corresponds to the stage when the H-burning shell crosses the hydrogen abundance discontinuity left over by
the convective envelope at the completion of the FDU. During this crossing, the luminosity of RGB models
drops temporarily, then starts again to increase after the hydrogen shell has moved beyond the discontinuity.
As a result, there is a narrow luminosity range along the RGB that is crossed three times by models. This produces 
a {\sl bump} (a local maximum) in the luminosity functions of old stellar populations
\citep[see, e.g.,][and references therein]{th, cs:13}.

As shown by Fig~\ref{figbump}, models predict than an increase of $Y$ increases the
bump brightness, and causes a smaller luminosity excursion.
The first effect is due to the lower envelope opacity of He-rich models, which shifts to more external layers
the discontinuity of the H abundance, because of a shallower surface convection.
A shallower convective envelope has also the important consequence that
the amount of He dredged to the surface by the FDU decreases when increasing
the initial He abundance, as shown in the upper panel of Fig.~\ref{figtip}. 

The second effect (smaller luminosity excursion) is caused by the narrower  
jump of the H abundance at the discontinuity in He-enhanced models, because of the lower H-abundance in the envelope.
As a consequence the surface luminosity is less affected when the H-burning shell
crosses this chemical discontinuity \citep[see,][for a detailed discussion on this issue]{csb:02}. 

The RGB bump vanishes when the initial $Y$ is larger than $Y_{\mathrm crit}=0.35$.
The exact value of $Y_{\mathrm crit}$ does depend on the
isochrone age and metallicity. It increases when the age decreases at fixed metallicity, and decreases when metallicity
decreases at fixed age.
For instance, at $Z$=0.002 the critical abundance $Y_{\mathrm crit}$ is equal 
to 0.32 at an age t=12.5~Gyr, whereas $Y_{\mathrm crit}=0.34$ at 12~Gyr \citep[see the discussion in][]{cs:13}.
On the other hand, at t=12~Gyr $Y_{\mathrm crit}$=0.28 when $Z$=0.0003.

Observationally, the impact of a He abundance range on the RGB bump brightness in Galactic GCs
has been discussed by \cite{bragaglia:10}. The effect on the RGB bump luminosity excursion
has been used by \cite{nataf:11} to interpret the small number of RGB
bump stars in the Galactic bulge: An oberved  number smaller than predictions of He-normal
stellar models has been considered as a proof that bulge stellar populations are He-enhanced.

The TRGB brightness is also affected by the initial He abundance. 
For a given value of the initial mass of RGB models, an increase of the initial He content decreases the TRGB brightness
(see Fig.~\ref{figtip}). The
reason is that He-rich models are hotter at the end of the MS, and 
develop a lower level of electron degeneracy in the He-core. At the same time, as a consequence of the higher H-burning
efficiency, the He-core mass grows at a  faster rate. Both effects acts in a way that the
thermal conditions required for the ignition of the 
$3\alpha$ reaction, are attained {\sl earlier}, with 
a smaller He-core mass, as shown in Fig.~\ref{figtip}. 
Due to the existence of a {\sl core mass-luminosity}
relation for RGB models, the TRGB brightness decreases in He-rich, low-mass giants.
We also recall that, as a consequence of the reduction of the MS lifetime of He-rich models, the mass of stars at the TRGB is
expected to be significantly smaller in He-enhanced stellar populations, when age is kept constant and if the efficiency of mass
loss during the RGB does not depend on the initial He content (see Fig.~\ref{figtip}). 

The effect of enhancing $Y$ at fixed [Fe/H] on the properties shown in Fig.~\ref{figtip} is almost exactly the same
as for the case of constant $Z$. The reason is that the variations of $Z$ associated to the increase of $Y$
(reported in Fig.~\ref{figisohe}) are small
enough to have a negligible influence on the results of Fig.~\ref{figtip}.

Regarding the bump luminosity, a given enhancement of helium at fixed age and [Fe/H] makes the bump luminosity to
increase either more or less than the case of enhancement at fixed $Z$, depending on the actual value of $Y$.
This is due to the combination of two factors. First of all, compared to the case of constant $Z$, 
an increase of $Y$ at fixed [Fe/H] implies
a reduction of the initial metallicity, which makes the bump brighter at fixed initial mass.
However, at a given age the mass of the models evolving
along the RGB is also reduced compared to the case at constant $Z$, causing a decrease of the bump luminosity.

Regarding the effective temperature of RGB isochrones, $Y$ enhancements at constant [Fe/H] make the isochrones hotter compared to
the case of enhancement at fixed $Z$, due to the metallicity decrease necessary to keep [Fe/H] constant. 

%\begin{figure*}
%\centering
%\includegraphics[width=9.5cm]{fig_hbsynthe.eps}
%\vskip -0.3cm
%\caption{{\sl Upper panel}:  the predicted distribution in the optical $M_V, (B-V)_0$ CMD of HB stars belonging to distinct sub-populations with the same metallicity and two different initial He contents (see labels), obtained by assuming that each one of their RGB progenitors has lost on average the same amount of mass $\Delta{M}=0.1M_\odot$, and that their (gaussian) mass distributions have the same dispersion $\sigma=0.015M_\odot$. {\sl Middle panel}: as upper panel but for $\Delta{M}=0.15M_\odot$. {\sl Lower panel}: as upper panel but for $\Delta{M}=0.2M_\odot$. In each panel, it is also listed the value of the $HB_{type}$ parameter for each individual sub-population (see text for more details).}
%\label{hbsynthe}
%\end{figure*}

\begin{figure}
\centering
\includegraphics[width=9.0cm]{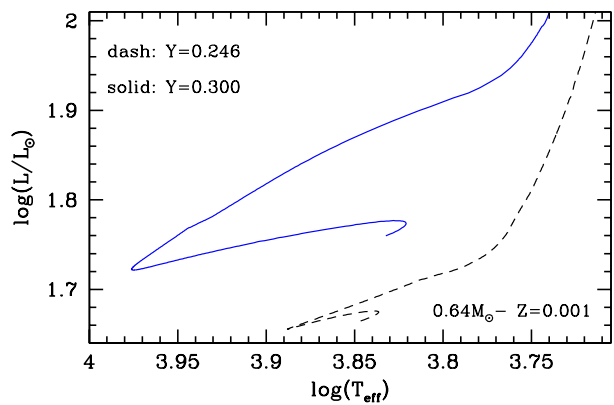}
\vskip -3cm
\caption{HRD of two HB models with the labelled mass and $Z$, but different initial He abundances (see labels).}
\label{fighbevohe}
\end{figure}

\begin{figure}
\centering
\includegraphics[width=9.0cm]{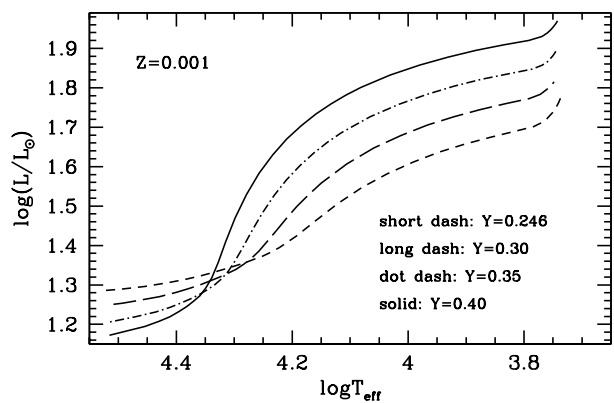}
\vskip -0.3cm
\caption{{\sl Upper panel}:  HRD of several ZAHBs with the same metallicity and different initial He abundances.
  The RGB progenitor has
  an age at the TRGB equal to $\sim12$~Gyr.}
\label{figzahb}
\end{figure}

\subsubsection{Core He-burning and He enhancement}
\label{sub:hb}

For a given population age, HB models for an enhanced
initial He abundance are typically bluer (hotter) in the HRD, compared to models for a normal initial He.
This is due to the combination of two effects
\citep[see also][for a detailed discussion on this topic]{dantona:02,caloi:05,cs:13}: 

\begin{itemize}

\item  As already mentioned, for a given age and $Z$, and assuming the same efficiency of mass loss along the RGB,
  the TRGB of a He-enhanced isochrone corresponds to models with  
  a smaller total mass at the TRGB and a smaller He-core mass than the He-normal counterpart.
  The first effect 
  prevails over the second one, even for variations of $Y$ as small as 0.01-0.02, hence the following HB
  phase in 
  He-enhanced populations will harbour models with a smaller total mass and a smaller $M_{env}/M_{tot}$ ratio.
  Furthermore, given that the  $T_{\mathrm eff}$ of a model on the ZAHB depends on this ratio
  --the lower the ratio, the hotter the ZAHB location-- the ZAHB of He-enhanced populations will be typically 
  bluer than the He-normal counterpart with the same age and $Z$;

\item As shown in Fig.~\ref{fighbevohe}, the larger the He abundance in the envelope of HB models at a given total mass,
   the more extended the blue loops in the HRD, which characterize the off-ZAHB evolution of He-rich models.

\end{itemize}

%Figure~\ref{hbsynthe} provides a plain evidence of this behaviour: it shows the synthetic distribution, in an optical photometric plane,
%of HB stars belonging to too distinct sub-populations, sharing the same metallicity and age (12.5~Gyr), but different initial He abundances
%(see labels). In the various panels, it is shown how the synthetic HB morphology changes when changing the average amount of mass lost by
%their RGB progenitors. The numerical simulations take into account of the fact that the RGB progenitors for each sub-populations have
%different initial masses: for the assumed metallicity $M_{pr}=0.84M_\odot$ for Y=0.248 (the He normal stellar populations), while for
%Y=0.30 $M_{pr}=0.76M_\odot$. Data in the figure disclose that, regardless of the adopted average amount of mass lost by the RGB progenitor
%He-enhanced HB stars are always bluer that their He-normal counterparts\footnote{In the figure it is also listed the value of the $HB_{type}$
%  parameter, that is defined as $HB_{type}=(B-R)/(B+V+R)$ where B, R, and V are the number of stars whose location is bluer or redder of the
%  RR Lyrae instability strip, or lie inside the strip. respectively. This parameter has been always used to characterize the HB morphology
%  in Galactic GCs.}.
%This figure also helps in explaining the existence of the very blue HB morphologies -- as well as the presence of extended HB blue tail --
%in GCs hosting He-enhanced  sub-populations such as the cases of NGC~2808 \citep[]{dantona:05,lee:05}, NGC6441 and NGC6388
%\citep[]{busso:07,dantona:08,tailo:17}.

A He-enhancement has an additional important implication for the global HB morphology.
As shown in Fig.~\ref{figzahb}, the ZAHB brightness is a strong function of the
initial He content: Typically, the higher the initial $Y$, the brighter the ZAHB. However, 
when $T_{\mathrm eff}$ is higher than $\sim 20,000$~K, the trend reverses.
This behaviour is a consequence of both the decrease of the He-core mass at the TRGB 
and the increased efficiency of shell H-burning in He-rich stars.
In models whose ZAHB location is cooler than $\sim20,000$~K, the second effect prevails, and He-enhanced models are brighter.
Above this $T_{\mathrm eff}$ threshold, due to the reduced envelope mass of the models, the H-burning shell efficiency is low, and
the decrease of the He-core mass causes a reduction of the ZAHB luminosity with increasing $Y$. 

%The fact that ZAHB He-rich stars with $T_{\mathrm eff}$ lower than  $\sim20000$~K are brighter than their He-normal counterpart,
%has the important consequence that the slope of the ZAHB in the H-R diagram (as well as in various CMDs) is strongly
%dependent on the (spread in the) initial
%He abundances of the various 
%sub-populations hosted within a given GC. This theoretical prediction provides a direct explanation for the existence of a tilted HB in GCs
%such as the cases of NGC~6388, NGC~6441 and NGC~1851\citep[see][and references therein]{caloi:07, busso:07, scp:08, tailo:17}.

\subsection{Isochrones for multiple stellar populations}

\begin{figure}
\centering
\includegraphics[width=7cm]{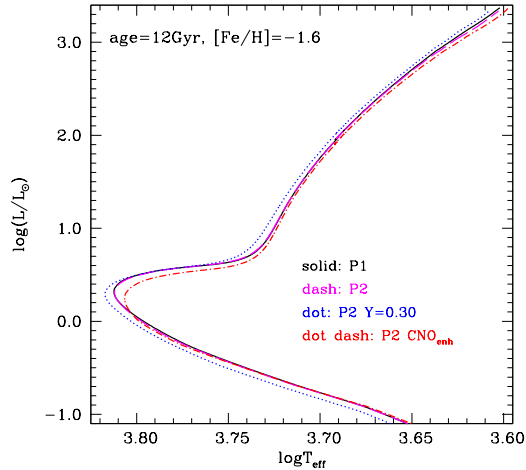}
\vskip -0.3cm
\caption{HRD of P1 and P2 (both with CNO sum like in P1 models, and CNO-enhanced) isochrones with the labelled age and [Fe/H].
 The P2 $Y$=0.30 isochrone has been calculated assuming the same CNO sum and $Z$
  of the P2 CNO-normal isochrone (it has therefore a slightly lower [Fe/H], by just a few 0.01~dex).}
\label{figisomp}
\end{figure}

Figure~\ref{figisomp} encapsulates the impact of several results of the previous sections on the HRD of cluster MPs.
We show here a comparison among isochrones from the MS (masses above $\sim$0.5$M_{\odot}$) 
to the RGB with the same age and [Fe/H], but different heavy element distributions,
namely P1, P2 with and without CNO enhancement, and P2
with He enhancement (this latter one has been calculated for the same $Z$ and CNO sum of the P2 CNO-normal composition).
Some important properties are worth mentioning:

\begin{itemize}

\item For P2 compositions at the same CNO of the P1 metal mixture, we expect that only variations of $Y$ affect the HRD.
  Higher $Y$ means hotter MS, TO (at fixed age) and RGB. The TO is only slightly fainter than P1 isochrones, the SGB essentially
  overlaps
  in luminosity with the P1 one;

\item An increase of the CNO sum affects the TO-SGB region of the isochrones.
 The TO becomes cooler and fainter, the SGB becomes fainter. The MS and RGB are unaffected;     

\item There is no need to calculate theoretical isochrones for P2 compositions, as long as the CNO sum in P2 stars is the same as in P1 objects.
  This allows us to employ standard $\alpha$-enhanced (or solar scaled) isochrones to model P2 populations, with either
  normal or enhanced He.
  
\end{itemize}

\section{Impact of abundance anticorrelations on theoretical spectral energy distributions}
\label{SED}

As already mentioned, the interpretation of photometric observations of cluster multiple populations requires
to asses the impact 
of the P2 chemical patterns not only on theoretical stellar evolution models and isochrones, but also
on the predicted SED of the models. 

%not only 
%theoretical evolutionary models and isochrones with the appropriate chemical patterns, but 
%also consistent model atmospheres and synthetic spectra.
%Once the effect of the abundance anticorrelations also on spectra and bolometric corrections is understood, 
%we can then employ photometry -- complementary to high-resolution spectroscopy -- to identify the various 
%subpopulations hosted by a cluster, quantify their differences, follow their evolutionary sequences in the CMD, 
%estimate the fraction of first to second generation stars, determine their radial profiles.
Early works which cross-correlated chemical abundances determined from spectroscopy, with photometry
of small samples of RGB stars in the Galactic GCs M4 and NGC6752, have 
found that the $(U-B)$ colour and the Str\"omgren $c_1$ index are able to separate the different populations
within the same cluster \citep{M4UBV, yong:08}.
The first theoretical investigation of the effect of P2 abundance patterns on stellar SEDs
was then performed by \citet{sswc:11} and later extended by \citet{cmpsf:13} and \citet{s19}.
More targeted studies focused on specific evolutionary phases or wavelength ranges have been published, e.g.,
by \citet{milone:12a}, \citet{milone:12c}, \citet{dotter}, \citet{nieder17}.

\begin{figure*}
\centering
\includegraphics[width=10.0cm]{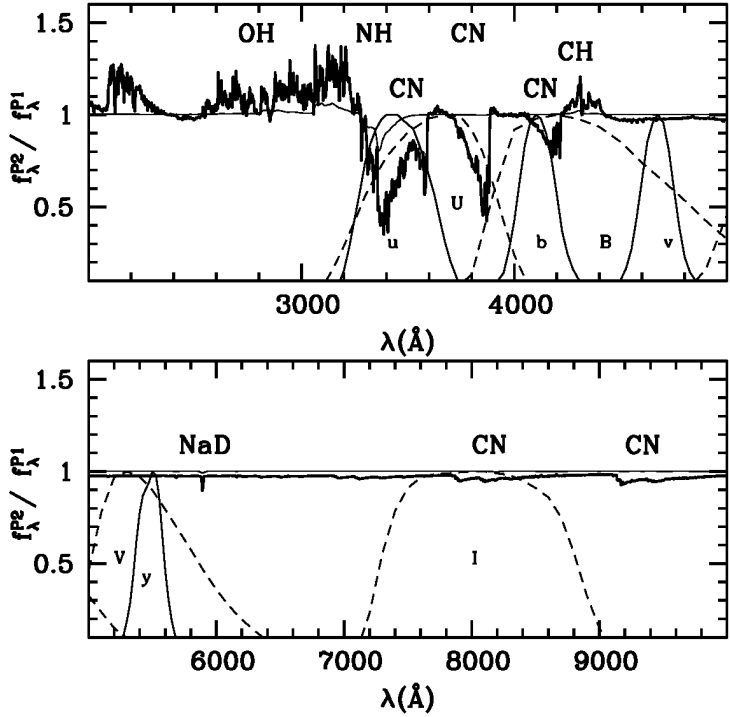}
\caption{{\sl Top panel}: Ratio of the fluxes calculated from P1 and P2 RGB model atmospheres 
  with $T_{\mathrm eff}$=4,500~K, surface gravity $\log(g)$=2 (cgs units -- thick solid line) and TO model atmospheres with
  $T_{\mathrm eff}$=6,131~K, $\log(g)$=4.5 (thin solid line), respectively. The wavelength range between
  2000 and 5000~\AA is displayed, together with the transmission curves of the $U$ and $B$ Johnson filters, plus the
  $u$, $b$, and $v$ Str\"omgren filters.
  The labelled molecules mainly contribute to the absorption in those wavelength ranges where P1 and P2 results are different.
  {\sl Bottom panel}: As the top panel, but for the wavelength range between 5,000 and 10,000~\AA, and the transmission curves of
  the $V$ and $I$ Johnson-Cousins filters, and the $y$ Str\"omgren filter.}
\label{figsedshort}
\end{figure*}

Figure~\ref{figsedshort} displays the ratio of the fluxes of two pairs of P1 and P2 SEDs computed for two
points along a 12~Gyr, [Fe/H]=$-$1.6, $Y$=0.246 isochrone. The first pair of P1-P2 SEDs is calculated along the RGB at  
$T_{\mathrm eff}$=4,500~K, and surface gravity $\log(g)$=2 (cgs units); the second pair corresponds to the
TO point, with $T_{\mathrm eff}$=6,131~K and $\log(g)$=4.5 \citep[calculations from ][]{s19}.
%The P1 and P2 compositions are described in Sect.~\ref{light}, and include also the Mg-Al anticorrelation
%\citep[from][]{cmpsf:13}.
The SEDs cover the wavelength range between 2,000 and 10,000~\AA, and 
the next Fig.~\ref{figsedlong} displays the continuation of the same SEDs up to 51,000~\AA.

The figures show very clearly the much 
stronger molecular NH and CN absorption bands between about 3,200 and 4,200 \AA\
in the P2 RGB SED, due to the higher N abundance. The CH molecular band at 4,400 \AA\ is less prominent in the P2 SED
due to the decreased carbon.
Below about 3,200 \AA\ the OH bands are weaker in the P2 SED, because of the lower oxygen abundance, while above 4,200 \AA\ the
differences between the two SEDs are very small.
Above 10,000 \AA\ the weaker CO bands in the P2 SED cause some sizable differences around 24,000 \AA\ and above
44,000 \AA.
In general differences decrease with increasing $T_{\mathrm eff}$: at the MS TO the two SEDs are basically identical,
apart from a small difference in a narrow wavelength range around 3,300 \AA\ due to NH, because the high temperatures
prevent the formation of CN and CH molecules.
Below about 4,000~K and at moderate-high
metallicities, molecular bands of TiO and ${\rm H_2O}$ also appear above 10,000~\AA, which are weaker in P2 SEDs due to the reduced 
abundance of oxygen.

\begin{figure}
\centering
\includegraphics[width=8.0cm]{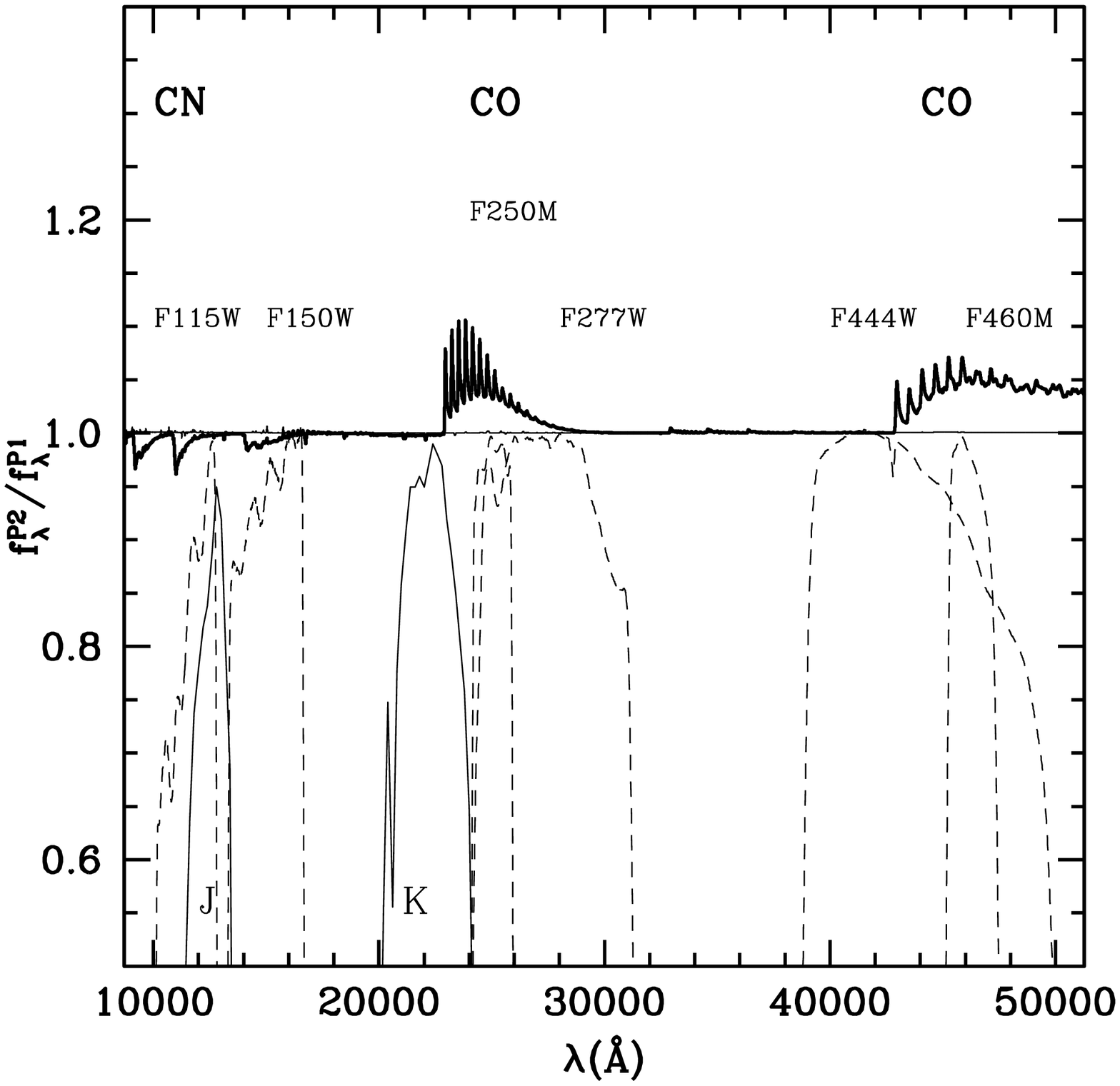}
\vskip -0.3cm
\caption{As Fig.~\ref{figsedshort}, but for the wavelength range between 10000 and  51000~\AA.
  The transmission curves of selected filters for the NIRCAM camera on board the James Webb Space Telescope (JWST),
  plus the $J$ and $K$ Bessell filters are also displayed.}
\label{figsedlong}
\end{figure}

Figures~\ref{figsedshort} and \ref{figsedlong} display also the transmission curves of representative photometric filters,
and show how below 10,000 \AA\ the Johnson $U$ and Str\"omgren $u$ filters are the most affected by the P2 chemical pattern,
for they cover
the wavelength range of prominent NH and CN molecular bands. The $B$ (in this wavelength range the opposite
effects of the CN and CH molecular bands almost compensate each other), $V$ and $I$ Johnson filters, as well as
the $b$ (only marginally affected by CN molecular absorption), $v$, $y$ Str\"omgren ones are much less affected, if at all.
When moving to the infrared, the $K$ fiter is marginally sensitive to a CO molecular band, while the filters
$F250N$, $F277W$, $F444W$ and $F460M$ are much more sensitive to the P2 chemical pattern.
Below 4,000~K and at higher [Fe/H] the appearance of ${\rm H_2 O}$ and TiO molecular bands affect all filters shown in
Fig.~\ref{figsedlong}.

\begin{figure*}
\centering
\includegraphics[width=8.0cm]{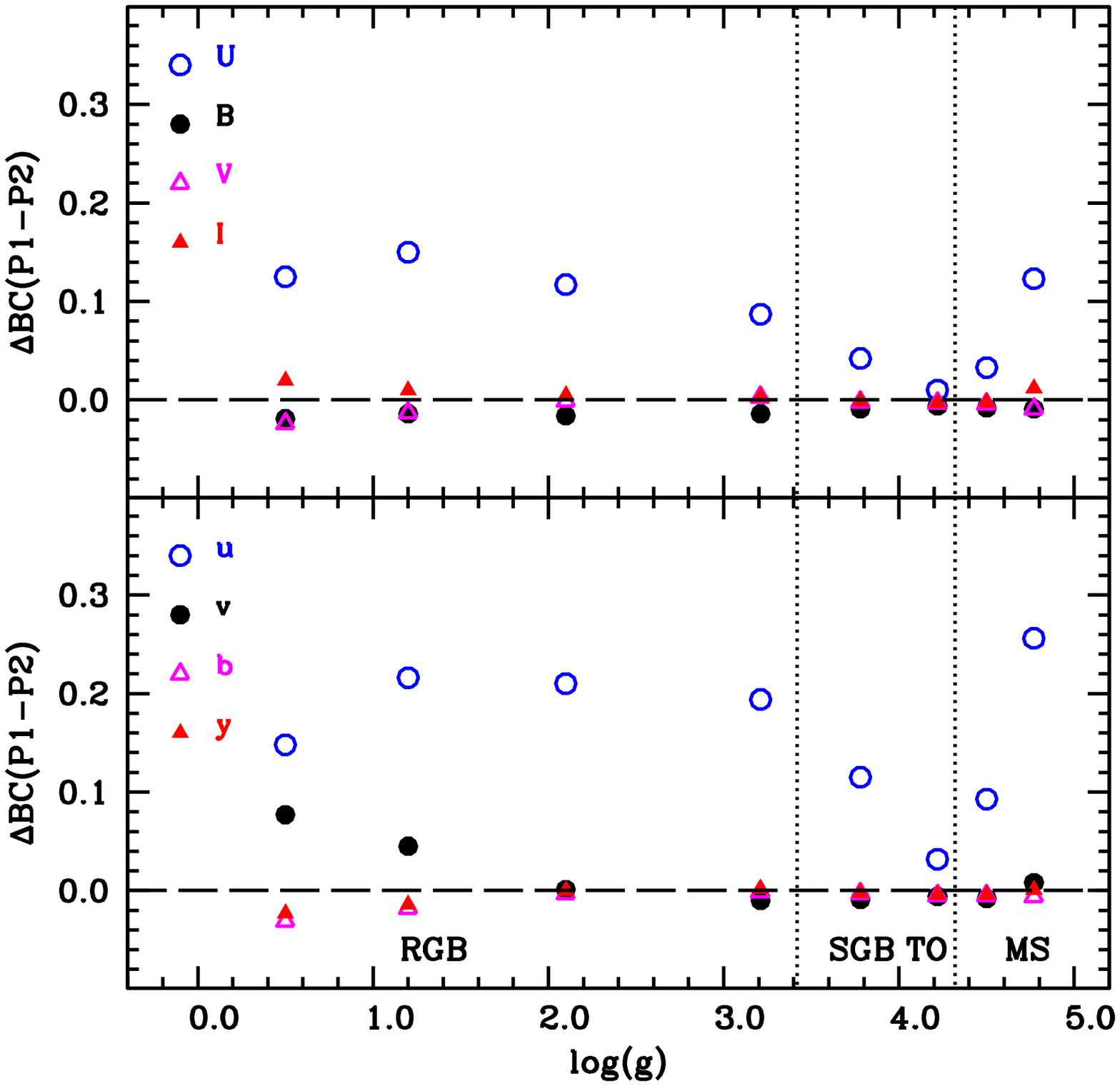}
\caption{Difference ($\Delta$(BC)) between BCs calculated
  for P1 and P2 compositions (P1-P2), at selected points along a 12 Gyr isochrone with [Fe/H]=$-$1.6 and $Y$=0.246.
  $\Delta$(BC) is plotted as a function of the surface gravity, which decreases steadily from the lower MS to the TRGB.
  We also label the evolutionary phases
  --MS, TO, SGB and RGB-- corresponding to the selected isochrone points.
  The top panel displays differences for Johnson-Cousins filters, the bottom panel shows results for Str\"omgren filters.}
\label{figdeltaBC}
\end{figure*}

Figure~\ref{figdeltaBC} shows
how the differences in the theoretical P1 and P2 SEDs translate to variations of the predicted magnitudes in selected
photometric filters. We display 
differences of the bolometric corrections (BCs) calculated for the
Johnson-Cousins and Str\"omgren filters of Fig.~\ref{figsedshort},
taken at selected points along a 12~Gyr isochrone with [Fe/H]=$-$1.6 and $Y$=0.246\footnote{We recall that, as long as the
  CNO sum (and initial $Y$) is the same, P1 and P2 theoretical isochrones are identical.}.
Only the $U$ and $u$ magnitudes change drastically between P1 and P2 composition, as expected.
Differences of BCs between P1 and P2 SEDs
are basically zero at the TO --the hottest point along the isochrone-- and increase when moving towards the lower MS and
along the RGB. 
On the RGB the values of $\Delta$BC reach a maximum at $T_{\mathrm eff}\sim$4,500~K, and decrease slightly towards lower
effective temperatures (corresponding to higher bolometric luminosities).
The effect on the BCs for the infrared filters of Fig.~\ref{figsedlong} is quantitatively smaller,
even for the $F250N$, $F277W$, $F444W$ and $F460M$ filters centred on CO molecular bands.
If we neglect the Mg-Al anticorrelation in the SED calculations or we increase $Y$, the effect on both SED and BCs is very minor
\citep[see, e.g.,][]{gc07, cmpsf:13, He18}.

\begin{figure*}
\centering
\includegraphics[width=8.0cm]{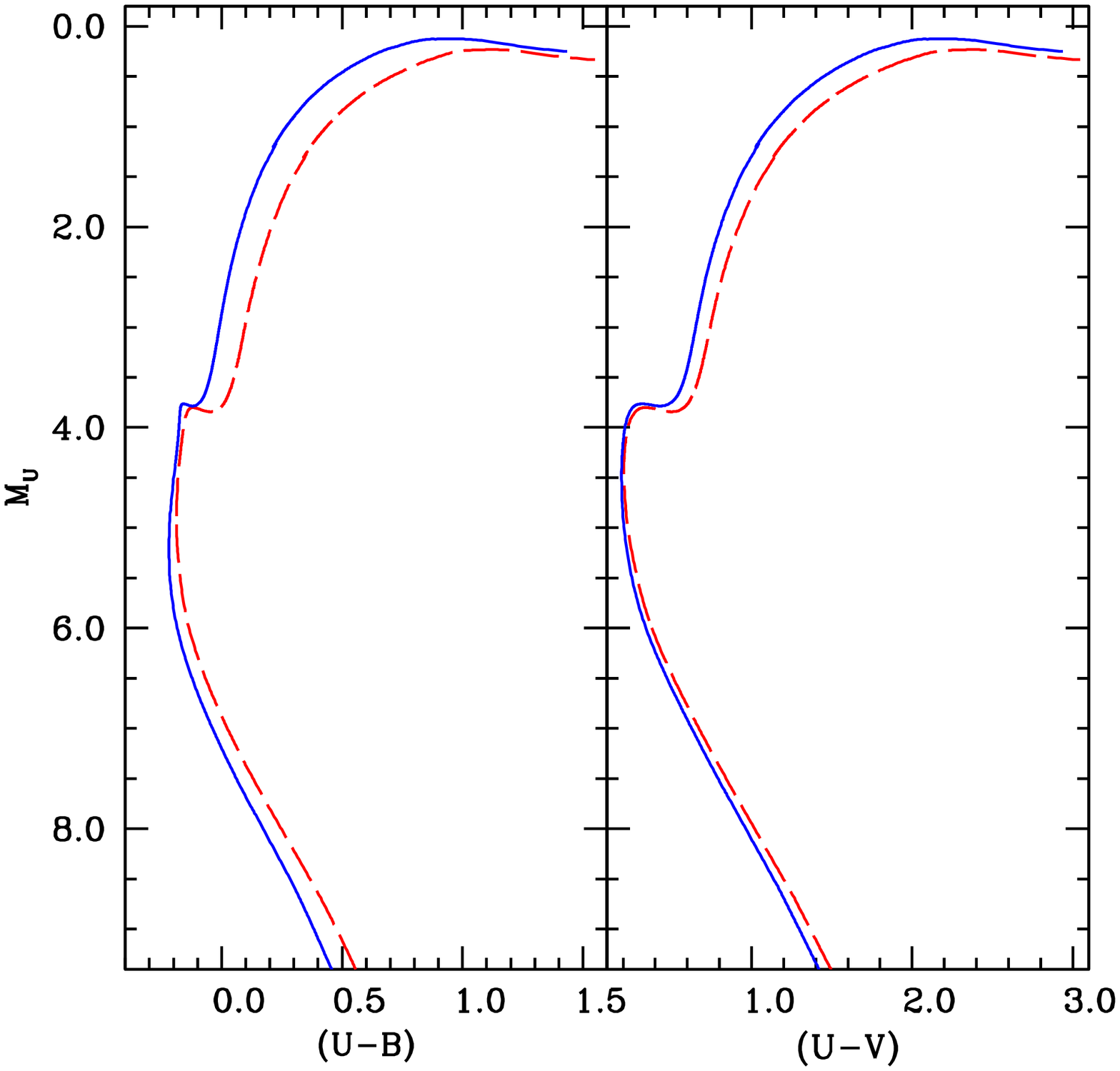}
\caption{$M_U$-$(U-B)$ and $M_U$-$(U-V)$ CMDs for 12~Gyr old P1 (solid lines) and P2 (dashed lines)
  isochrones,  with [Fe/H]=$-$1.6 and $Y$=0.246.}
\label{figisoUV}
\end{figure*}

Figure~\ref{figisoUV} shows the 12~Gyr old [Fe/H]=$-$1.6, $Y$=0.246 isochrones for P1 and P2 compositions in the $UV$ and $UB$
colour-magnitude diagram (CMD). They are clearly separated along MS and RGB -- P2 models being redder
in $(U-B)$ and $(U-V)$-- whilst they converge around the TO,
where the BCs are essentially the same for P1 and P2 SEDs.
This separation is due to the effect of CN and NH molecular absorption in the $U$ band, whilst the $B$ and $V$ bands are
essentially unaffected by the P2 abundance anticorrelations.

\begin{figure*}
\centering
\includegraphics[width=10.0cm]{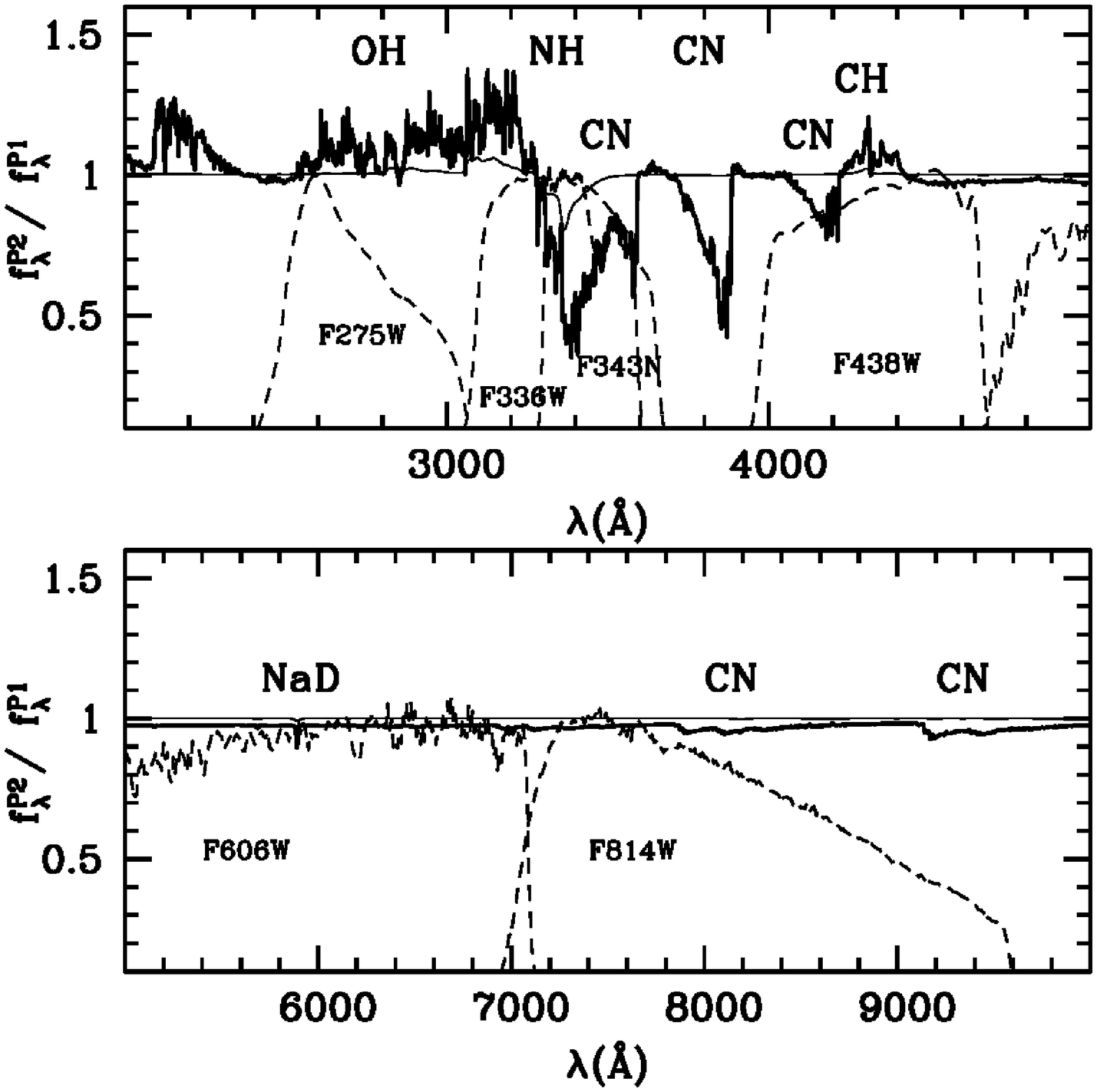}
\caption{As Fig.~\ref{figsedshort}, but displaying the transmission curves of selected photometric filters of the WFC3 (top panel)
  and ACS (bottom panel) cameras on board HST.}
\label{figsedshortb}
\end{figure*}

Passbands of selected filters of the WFC3 and ACS cameras on board the Hubble Space Telescope (HST), and widely used to
disentangle cluster MPs --like for example in the {\sl UV Legacy Survey of Galactic Globular Clusters}-- 
are displayed in Fig.~\ref{figsedshortb}. We can see clearly how the $F336W$ and $F343N$ passbands are very sensitive to
the nitrogen abundance, while the effects of oxygen variations on the $F275W$ filter,
and the C-N anticorrelation on the $F438W$ filter are less substantial.

On the quantitative side, the size of these effects depend obviously on the exact values of C and N abundances in the P2
SED calculations (in these tests we employed the values in Fig.~\ref{anticorr}), and the 
the metallicity $Z$, which affects also the $T_{\mathrm eff}$ of the underlying theoretical isochrones \citep[see, e.g.,][]{He18}.
If we combine the results about the effect of P2 compositions (at fixed $Z$ and age)
on both theoretical isochrones and SEDs, we can draw the following
general conclusions about photometric diagnostics of cluster MPs:

\begin{itemize}

\item Spreads and/or splittings of both MS and RGB in optical CMDs (Johnson-Cousins $BVI$ or HST equivalent filters)
  are caused by variations of the initial helium content $Y$ between P1 and P2 stars.

\item Spreads and/or splittings of the TO-SGB region in optical CMDs are caused by variations of
  the CNO sum.

\item CMDs which employ colours like, e.g., $(U-B)$, $(U-V)$, $(F336W-F606W)$, $(F336W-F814W)$, $(u-y)$ --made of a
  passband sensitive
  to the CN anticorrelation, and a passband unaffected by CN molecular bands-- will show spreads and/or splittings along MS, SGB
  and RGB, with a minimal effect around the TO. Colours like $(F275W-F814W)$ and $(F438W-F814W)$ are sensitive, to smaller degrees
  (but this also depends on $Z$),
  to oxygen (through OH molecular absorption) and carbon (through CH molecular absorption) depletion, respectively. 
  In addition, these CMDs will show the additional effect of variations of $Y$ (if present), which alter the 
  $T_{\mathrm eff}$ of the MS, TO and RGB, like in case of the optical CMDs. 
\end{itemize}

\section{Colour magnitude diagrams}
\label{chrom}

As discussed at the end of the previous section, CMDs with appropriately chosen colours including a passband (typically
in the UV) sensitive to the CN anticorrelation, are able to disclose the presence of MPs in clusters. They show up as
broadenings or multimodalities in the colour distributions of RGB and MS stars.
The use of optical colours can still reveal MPs along MS and RGB, because a range of initial He abundances causes again 
colour spreads and multimodalities.
In general, whenever the colour distribution at fixed magnitude along a given evolutionary phase --the large majority of the
investigations are based on the RGB, that is
bright and well populated in old- and intermediate-age populations-- is broader than what expected from photometric errors only,
MPs are present in the cluster. If bi- or multimodality is found in these colour distributions, P1 and P2 (one or
several P2 subpopulations) can be separated, and their fractions of the total cluster population can be estimated. 
Moreover, once disentangled, their radial distributions within a cluster can be determined and compared.

Just to give a few examples, \citet{piotto:07} employed an ACS 
$F814W$-$(F475W-F606W)$ CMD to detect multiple sequences along the MS -- hence a multimodal He abundance distribution--
in the Galactic GC NGC~2808.
More recently, \citet{milone:12a} and \citet{p13} made use of several
WFC3 and ACS CMDs, employing 
various filters from the UV to the optical. By accounting for the response of various colour combinations to variations
of C, N and He, these authors determined, by an iterative procedure, the presence of abundance spreads 
along the MS and RGB of of the Galactic GCs 47~Tuc \citep{milone:12a} and NGC~288 \citep{piotto:07}.
\citet{milone:12a} detected MPs also along the SGB and HB of 47~Tuc.

The theoretical justification for this procedure is that different colours respond differently
to variations of these three elements (for example an optical colour like $(F606W-F814W)$ is affected only by
variations of $T_{\mathrm eff}$ hence $Y$, whilst e.g., $(F336W-F814W)$ is sensitive to N).
An analogous technique was applied by \citet{He18} and \citet{lagioiab}
to determine He abundance difference between P1 and P2 populations
(disentangled by means of the so-called {\sl chromosome maps} discussed below) in 57 Galactic GCs and
four intermediate-age and old Small Magellanic Cloud clusters,
considering multiple colour differences at a given reference magnitude along the cluster RGBs.

Analogous technique was employed by \citet{milone:14}, but applied to the lower MS of the Galactic GC M4
and using different sets of WFC3
and ACS filters, including the IR filters $F110W$ and $F160W$ of the WFC3 camera. The CMD $F160W$-($F110W-F160W$) was found 
to be especially sensitive to MPs at these low $T_{\mathrm eff}$ (below about 4,000~K)
and [Fe/H]$\sim -$1, because of the effect of ${\rm H_2O}$ molecular bands on the theoretical SEDs. The observed
spread of the lower MS in this colour is effectively tracing the presence of the O-Na anticorrelation.
We also note that this detection of MPs in fully convective, lower MS stars, has provided strong evidence
of their primordial origin.

\subsection{Pseudocolours and chromosome maps}

A slightly different approach devised to enhance the photometric separation between P1 and P2 stars 
envisages the use of appropriate combinations of colours, that we will refer to interchangeably as 
colour indices or pseudocolours. The basic idea is to use a combination of, for example, 
two colours, one of which increases whilst the other one decreases when moving from P1 to P2 compositions.
Their combination would therefore enhance the disentangling power, compared to just one colour. In same cases the
combination 
is chosen as to erase the dependence on $T_{\mathrm eff}$ of a colour sensitive to P2 chemistry along the RGB,
to produce almost vertical sequences in a suitable magnitude-pseudocolour diagram. This facilitates the identification of 
broadenings/multimodalities caused by MPs.

\begin{figure}
\centering
\includegraphics[width=6.0cm]{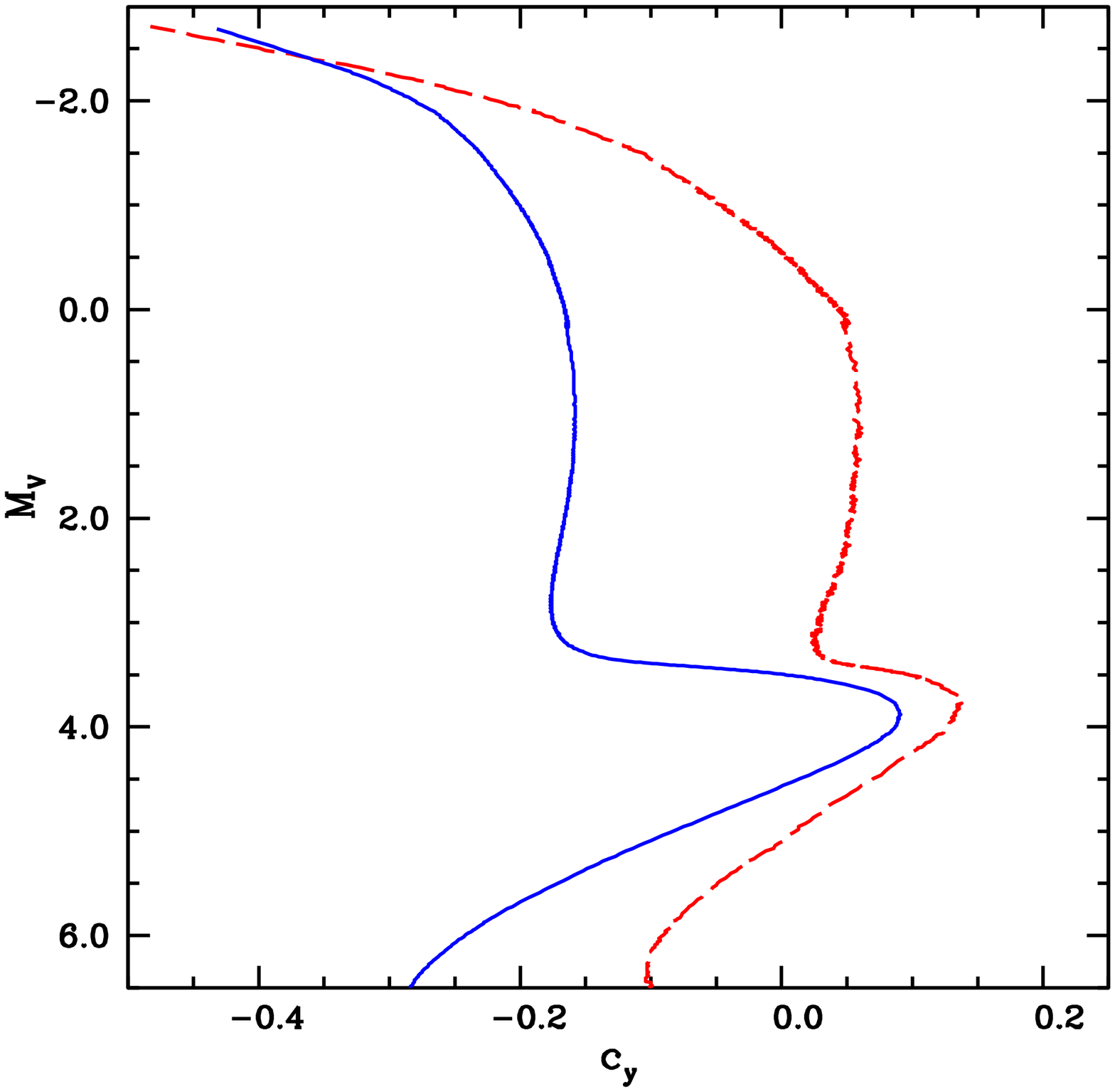}
\includegraphics[width=6.0cm]{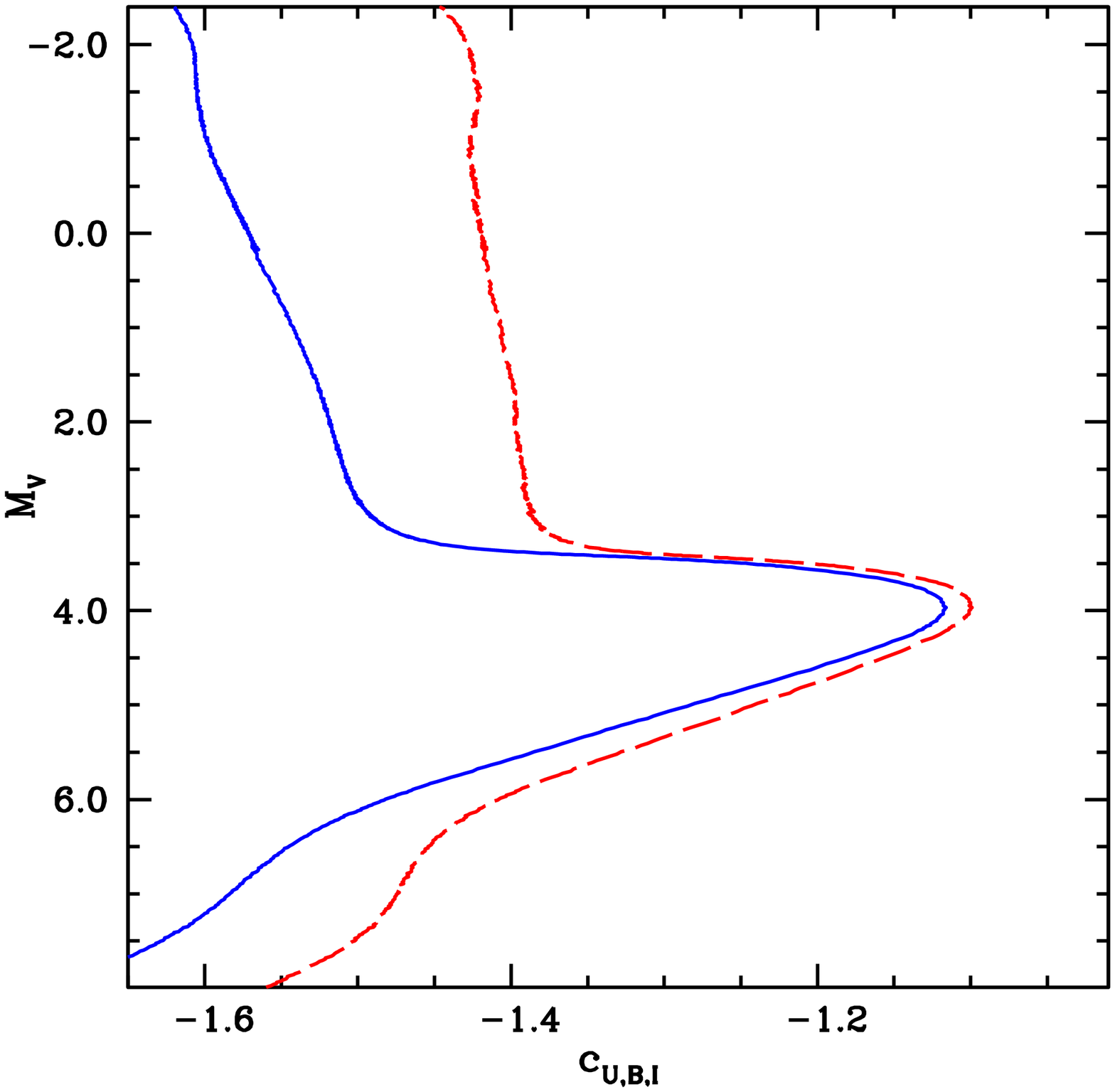}
\caption{{\sl Top panel}: A pair of 12~Gyr, [Fe/H]=$-$1.6, $Y$=0.246 isochrones, for both the P1 (solid line) and
  P2 (dashed line) compositions of Sect.~\ref{sec:scenario}, displayed in the $M_V$-$c_y$ diagram (see text for details).
  {\sl Bottom panel}: As the top panel, but for the $M_V$-$c_{U,B,I}$ diagram (see text for details).}
\label{pseudo1}
\end{figure}

An early example can be found in \citet{grundahl}, who found a correlation between the pseudocolour $c_1=(u-v)-(v-b)$
and the nitrogen abundances measured in a sample of RGB stars in the Galactic GC NGC~6752.
\citet{yong:08} later introduced the Str\"omgren index $c_y=c_1-(b-y)$, 
which tracks $c_1$, but removes much of its temperature sensitivity, as shown in Fig.~\ref{pseudo1}.

As a result, $V$-$c_y$ CMDs of Galactic GCs display an almost vertical RGB at luminosities 
lower than the RGB bump (see Fig.~\ref{pseudo1}), as shown empirically by \citet{yong:08}, who also derived a tight correlation 
between ${c_y}$ and spectroscopic N abundances aalong the RGB.
Theoretical isochrones displayed in the $M_V$-$c_y$ diagram agree with this empirical evidence. The sensitivity
to N comes mainly through the filter $u$, as discussed in the previous section.
P2 isochrones have redder $c_y$ colours than P1 ones, as shown in Fig.~\ref{pseudo1},
and an increase of the initial He abundance shifts the RGBs of P2 isochrones further away from the P1 ones \footnote{\citet{leejw}
has introduced the $cn_{JWL}$ 
index, defined as $cn_{JWL}=JWL39-Ca_{new}$, where $JWL39$ and $Ca_{new}$ are two fiters in the wavelength range between 3800 and
4050 \AA. The $cn_{JWL}$ index is sensitive to the absorption of the CN band at 3883 \AA, and the 
$V$-$cn_{JWL}$ diagram is very similar to the $V$-$c_y$ counterpart, but with a better MP resolving power.}

\citet{delta} have proposed instead the Str\"omgren index $\delta_4=(u-v)-(b-y)$. Calculations of P1 and P2 isochrones
show that in the $M_y$-$\delta_4$ diagram, P2 isochrones are systematically shifted to higher values of $\delta_4$ compared to
P1 ones, both along MS and RGB. Moreover, isochrones are almost insensitive to the initial $Y$, hence any colour spread 
observed in this CMD depends only on the range of N abundances\footnote{The $\delta_4$index is also almost 
insensitive to reddening, given that $E(\delta_4) \sim 0.01 E(B-V)$.}.

Moving to the Johnson-Cousins system, \citet{milone:12a} employed the $(U+B-I)$ pseudocolour to disentangle MPs along
the red HB of 47~Tuc in a $B$-$(U+B-I)$ diagram, whilst 
\citet{sumo} introduced the pseudocolour     
$C_{U,B,I} = (U-B)-(B-I)$, which displays a behaviour very similar to $c_y$, as shown in
Fig.~\ref{pseudo1}. This $C_{U,B,I}$ index is sensitive to the nitrogen 
abundance through the $U$ filter (like the case of the $(U+B-I)$ pseudocolour)
and as for $c_y$, an increase of the initial He abundance shifts
the RGBs of P2 isochrones further away from the P1 ones.
\citet{sumo} have employed $M_V$-$C_{U, B, I}$ diagrams of 23 Galactic GCs to detect the presence
of MPs by analyzing
the $c_{U,B,I}$ spread along the RGB.

The pseudocolours $C_{F343N, F438W, F814W}= (F343N-F438W)-(F438W-F814)$ and $C_{F336W, F438W, F814W}=(F336-F438W)-(F438W-F814)$ have been
introduced by \citet{nieder17} and \citet{martocchia17} to detect MPs along the RGBs of massive clusters in the Magellanic Clouds,
by comparing the observed width with that expected from photometric errors only (once differential reddening, if any, has been
corrected for).
In each of these combinations, the first colour is sensitive to N enhancements,
and gets redder for P2 compositions, whilst the second colour has a mild sensitivity to carbon depletions, and gets
bluer for P2 compositions.
These pseudocolours thus enhance the separation of the various cluster populations compared to using just one N-sensitive
colour.

\begin{figure*}
\centering
\includegraphics[width=10.0cm]{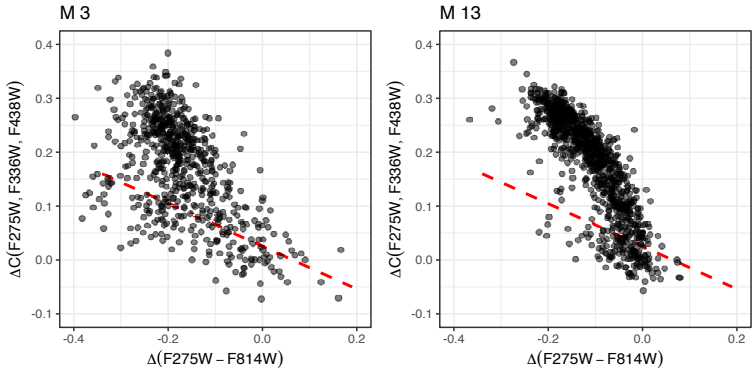}
\caption{Chromosome maps for the two Galactic GCs M~3 and M~13.
%  The arrows in the M3 map 
%  display the effect of varying the abundance of nitrogen and helium, respectively.
  The dashed slanted line in each panel shows the boundary between P1 and P2 stars (see text for details).}
\label{chromo}
\end{figure*}

A new and different
diagnostic tool which makes use of suitable colours and pseudocolours was presented by \citet{m17}, as the culmination of the
HST {\sl UV Legacy Survey of Galactic Globular Clusters}. This tool, applied to 57 clusters of the survey,
has been named {\sl chromosome map} and is based on the combination of a $F814W$-$(F275W-F814W)$ CMD
and a $F814W$-$C_{F275W, F336W, F438W}$ diagram, where
$C_{F275W, F336W, F438W}=(F275W-F336W)-(F336W-F438W)$\footnote{See also 
  \citet{milone:15} for a first discussion of this map for the Galactic GC NGC~2808, and alternative versions
  of chromosome maps.
  \citet{z19} has very recently used another version of chromosome maps to study the Galactic GC NGC~2419,
  employing the pseudocolour
  $C_{F275W, F343N, F438W}=(F275W-F343N)-(F343N-F438W)$ and the colour $(F438W-F814W)$. The construction and general properties of this
  map are the same as for \citet{m17} maps. For the same cluster, \citet{larsen19} have devised an analogous type of
  chromosome map, but based on the $(F438W-F814W)$ and $(F336W-F343N)$ pair of colours.}.

Given a cluster members' photometry (corrected for differential reddening, if any),
red and blue fiducial lines of the RGB in each of these two diagrams are calculated by
considering the values of the 4th and the 96th percentile of the $(F275W-F814W)$ and $C_{F275W, F336W, F438W}$ distributions
in various magnitude bins.
On theoretical grounds, in both the $F814W$-$(F275W-F814W)$ HRD and
$F814W$-$C_{F275W, F336W, F F438W}$ diagram, P2 stars are expected to be
located at blue side of the observed RGB.
As a second step, the widths of the RGB in the $(F275W-F814W)$ colour ($W_{F275W,F814W}$) and in the $C_{F275W, F33W, F438W}$
pseudocolour ($W_{C\ F275W,F336W,F438W}$) two $F814W$ magnitudes above the TO, are calculated by 
taking the colour and pseudocolor differences between the red and blue fiducials at the reference $F814W$ magnitude.
All clusters studied by \citet{m17} have $W_{F275W,F814W}$ and $W_{C\ F275W,F336W,F438W}$ larger than zero, even after
subtracting the effect of photometric errors, implying that MPs are ubiquitous in their sample.
Finally, for each star the following quantities are calculated:

\begin{equation}
\Delta _{F275W,F814W}= W_{F275W,F814W} \frac{X-X_{\rm R}}{X_{\rm R}-X_{\rm B}}
\end{equation}

\begin{eqnarray}
&&{\Delta _{C\ F275W,F336W,F438W}}\nonumber\\
&&{\quad= W_{C\ F275W,F336W,F438W} \frac{Y_{\rm R}-Y}{Y_{\rm R}-Y_{\rm B}},}
\end{eqnarray}

where X=$(F275W-F814W)$ and Y=$C_{F275W,F336W,F438W}$ are measured, R and B
correspond to the values for the red and blue fiducials at the star $F814W$ magnitude.

With these definitions, $\Delta_{F275W, F814W}$=0 and $\Delta_{C\ F275W, F336W, F438W}$=0 
correspond to stars lying on the red fiducial
lines --P1 stars, which will be spread around these coordinates due to the photometric error-- 
and $\Delta$ values different from zero denote colour and pseudocolour distances (defined as positive
for $C_{F275W,F336W,F438W}$, and negative for $(F275W-F814W)$) from such lines.
The values of these distances are normalized to the distance between the two fiducials 
taken at the reference $F814W$ level above the TO.

Figure~\ref{chromo} displays, as an example, chromosome maps of the two Galactic GCs M~3 and M~13, which have essentially the
same age
and [Fe/H] \citep[see, e.g.,][]{sw}.
%The arrows display the effect of varying the abundance of nitrogen and helium,
%the main elements affecting the position of the cluster RGB stars in this diagram
%\citep[a detailed analysis of the effect of varying the individual abundances of various elements has been
%  performed by][]{He18,lsb18}.
%of the cluster RGB stars in this diagram.

%The six maps calculated for each cluster differ regarding the magnitude level employed to normalize 
%the $\Delta$ values. 
%Let's consider first the map with the standard normalization taken 2 $F814$ magnitudes above the TO, together with 
%the arrows that display the effect of varying the abundance of nitrogen and helium, the main elements affecting the position
%of the cluster RGB stars in this diagram.

The distribution of stars describes a curve extended towards increasing $\Delta_{C\ F275W, F336W, F438W}$ and
decreasing $\Delta_{F275W, F814W}$,
implying the presence of RGB stars bluer than the P1 population in both $(F275W-F814W)$ and $C_{F275W,F336W,F438W}$.
The relative narrowness of the sequence also indicates that the shifts in $(F275W-F814W)$ and $C_{F275W,F336W,F438W}$ are
correlated.
An increase of N moves the position of stars almost vertically towards higher $\Delta_{C\ F275W, F336W, F438W}$,
whilst an increase of He ($(F275W-F814W)$ is strongly sensitive to $T_{\mathrm eff}$) and a decrease of O shift stars 
roughly horizontally towards lower $\Delta_{F275W, F814W}$.
At low- and intermediate metallicities like the case of these two clusters, 
$(F275W-F814W)$ is very weakly sensitive to the oxygen 
abundance \citep{lsb18}, and 
the shape of the cluster chromosome maps in Fig.~\ref{chromo} shows clearly that P2 stars are also enriched in He.

In general, the slope of the curve depends on how the increase of He and decrease of O
trace the increase of N, and this may
vary from cluster to cluster \citep[a detailed analysis of the effect of varying the individual abundances
of various elements can be found in][]{He18,lsb18}.
%If we compare the chromosome map of M13 with the one (calculated with the same normalization) of M3 (roughly same
%age and same [Fe/H]), the slope is not the same: 
%This can be seen by considering the relative position between the custer stars and the reference open rectangle located above
%$\Delta_{C\ F275W, F336W, F438W}$=0.2.

\begin{figure*}
\centering
\includegraphics[width=10.0cm]{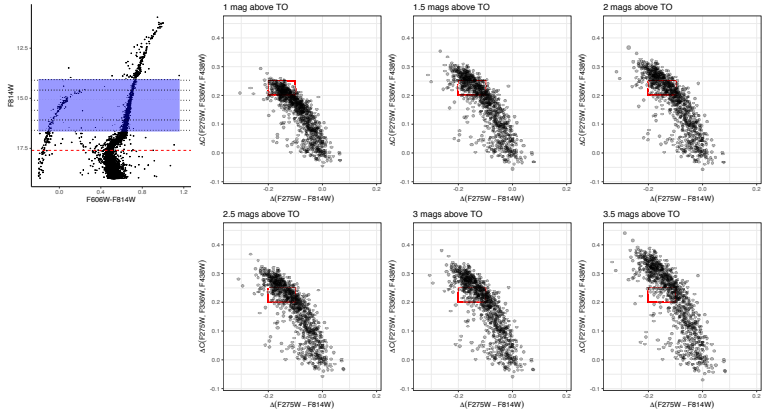}
\caption{{\sl Top panel}: Optical CMD, and 6 chromosome maps of the Galactic GC M~13, obtained by normalizing
  the values of the $\Delta$ quantities at different $F814$ levels, as labelled and
  marked in the optical CMD. The red open rectangle in the chromosome maps encloses the same range of
  $\Delta$ coordinates in all panels (see text for details).}
%  The arrows 
%  display the effect of varying the abundance of nitrogen and helium, respectively.
%  The dashed slanted line in each panel shows the boundary between P1 and P2 stars as defined by \citet{m17}.
% {\sl Bottom panel}: As the top panel, but for the Galactic GC M3.}
\label{chromob}
\end{figure*}

The chromosome maps of the majority of the GCs investigated show well separated groups of stars,
corresponding to P1 (clustered around the origin of the $\Delta$ coordinates) and P2 populations. In some cases
the P2 population consists of a few subgroups centred around discrete values of $\Delta_{C\ F275W, F336W, F438W}$.
Only for a handful of clusters there is no clear separation, between P1 and P2 along their chromosome map, which shows a
continuous number distribution along the full range of $\Delta_{C\ F275W, F336W, F438W}$.

Another important point to notice is that the majority of clusters in \citet{m17} sample display another sequence extending
from $\Delta_{F275W, F814W}$=0 and $\Delta_{C\ F275W, F336W, F438W}$=0 (the expected location of P1 stars) towards negative values
of $\Delta_{F275W, F814W}$, but with only a small increase of $\Delta_{C\ F275W, F336W, F438W}$. This can be seen in the map
of M~3 map 
(the sequence below the dashed line), but is virtually missing in M~13.
Correlations of spectroscopic abundance measurements for samples of RGB stars in NGC~2808 \citep{milone:15, cabrera:19}
and M~4 \citep{m17} with positions in the clusters' chromosome maps
show that this sequence hosts stars with [O/Fe] abundance ratios typical of P1 stars.  
The natural explanation for the existence of this sequence is therefore the presence of a range of
He abundances also in P1 stars, at least in a large fraction of GCs \citep{He18, lsb18}. An alternative explanation is
the presence of a small range of Fe abundances (Fe decreasing towards negative values of $\Delta_{F275W, F814W}$, because of
the higher $T_{\mathrm eff}$ of more Fe poor RGB stars) in P1 stars. 
\citet{marino:19} found hints of a range of [Fe/H] among P1 stars in NGC~3201, which shows an extended P1 sequence in
the chromosome map, but see also the discussion in \citet{muc15}.

On the other hand a range of Fe is not found by \citet{cabrera:19} along the extended P1 sequence of
NGC~2808.
A recent analysis by \citet{martins} has also excluded binaries \citep[see also][]{kamann}
and chromospheric activity as explanations for the observed extended P1 sequences.

In general, shape and extension of the sequences in the chromosome maps exhibit a great deal of variety.  
A number of clusters \citep[denoted as Type~II by][as opposed to Type~I clusters with the typical chromosome maps
just described]{m17},
like e.g., $\omega$~Cent, NGC~1851, NGC~6656, NGC~6934, even display additional P2 sequences (e.g. NGC~1851)
or an apparent split of both P1 and P2 sequences (e.g. NGC~6934), these additional features being typically located 
at larger $\Delta_{F275W, F814W}$ compared to the main ones. Some of these clusters are known to have a spread also
of [Fe/H] (e.g. $\omega$~Cent), others a spread of the CNO sum (e.g. NGC~1851); these further chemical peculiarities are
most likely the reason for the observed additional sequences in the chromosome maps.

A couple of final comments on these maps.
First of all, due to the effect of the metallicity on the colour and pseudocolour sensitivity to the MP abundance variations 
\citep[e.g.][]{He18}, comparison of cluster chromosome maps need to account for [Fe/H] differences.
Also, it is worth remarking that the exact shape and extension of the sequences depends
on the $F814W$ magnitude level at which the map is normalized. This can be seen in Fig.~\ref{chromob}, which displays
six different chromosome maps for M~13, with the normalization taken at several different levels above the TO.
The position of the reference open rectangle in each map relative to the P2 sequence, shows that the shape
(and the total range of the $\Delta$ coordinates) of the map depends on the normalization magnitude.
This is because, at fixed metallicity, the sensitivity of $(F275W-F814W)$ and 
$C_{F275W, F336W, F438W}$ to chemical abundance variations depend on $T_{\mathrm eff}$.
Especially in case of intermediate-age clusters, there is also the additional issue that  
the range of surface nitrogen abundances spanned by P1 and P2 stars generally changes 
along the RGB, as discussed below.
Both effects imply that the values of the widths 
$W_{F275W,F814W}$ and $W_{C\ F275W,F336W,F438W}$ do depend on the brightness level chosen for the normalization of the maps. 
This needs to be taken into account when
translating the observed shape and extensions of the chromosome maps to quantitative chemical abundance spreads.

\subsection{First dredge up and thermohaline mixing}

An important issue to consider when translating colour or pseudocolour ranges into initial chemical abundance spreads,
or when ranking the width of RGBs with respect to age and/or metallicity, 
is the evolutionary change of the surface abundances involved in the anticorrelations.
The large majority of the information obtained about MPs has come from RGB stars,
and we know that the FDU alters the surface abundance
of mainly nitrogen, and to a lesser degree carbon \citep[see, e.g.,][]{cs:13, s15}.
During the FDU the surface nitrogen increases (and carbon decreases) compared to the initial values, and
this increase depends on the initial abundance. The variation of
the surface abundances due to the FDU impacts the SEDs, hence the predicted colours
and pseudocolours sensitive to this element.
This is not relevant for old star clusters (especially the metal-poor ones), like the Galactic GCs,
because the impact of the FDU at these ages is minimal \citep[see, e.g.,][]{s15}, but 
the effect is more important in intermediate age clusters, where MPs have also been detected \citep{s20}.

Moreover, we know observationally \citep[see, e.g.,][]{gscb} that at the RGB bump an 
additional element transport mechanism starts to increase again the surface N
(and decrease C) with increasing luminosity.
Thermohaline mixing is generally considered to be responsible for these latter changes, although theoretical predictions
for the efficiency of this process in RGB stars are still uncertain
\citep[see, e.g.][]{chl10, lattanzio}.

Because of the FDU and later (presumably) thermohaline mixing,
the observed RGB widths in N-sensitive (and C-sensitive) colours and pseudocolours are determined
by both the initial nitrogen (and carbon) abundance spreads, and the variations induced
by these transport mechanisms.
P2 stars, with a higher initial N abundance, are predicted to be much less affected by 
FDU and thermohaline mixing, compared to P1 objects.
The reason is that during the FDU the convective envelope reaches
layers where the abundances of C and N attained the equilibrium values of the CN cycle during core H-burning.
The equilibrium abundance of N is typically higher (and
the C abundance lower) than standard P1 solar scaled or
$\alpha$-enhanced counterparts for a given total metallicity, hence
the FDU causes an increase of surface N (and a decrease of C).
However, when the initial metal mixture is N-enhanced (and carbon
depleted) the equilibrium abundance of N (and C) becomes
more comparable to the initial one, and the effect of the
FDU is much less appreciable, or even negligible.
The same is true for thermohaline mixing, which acts as to erase the gradients of N and C abundances between the base of
the convective envelope and inner layers where CN is at equilibrium.

\begin{figure}
\centering
\includegraphics[width=7.0cm]{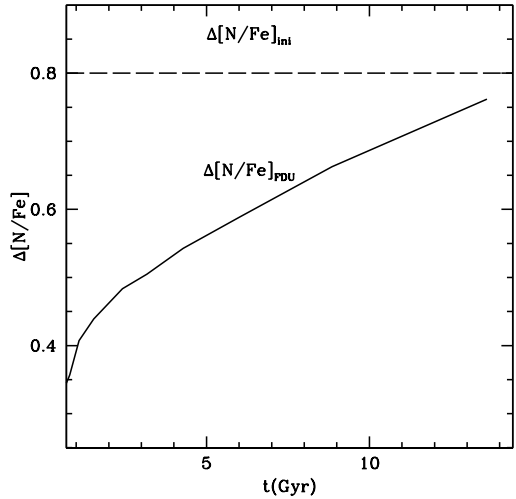}
\caption{Initial value ($\Delta {\rm [N/Fe]_{ini}}$=0.8 -- dashed line)
of the surface [N/Fe] difference for a set of bimodal populations
with [Fe/H]=$-$1.3 and various ages $t$,
together with the corresponding difference $\Delta {\rm [N/Fe]_{FDU}}$
in the surface abundances after the FDU (solid line).}
\label{DNfdu}
\end{figure}

Reliable predictions for the effect of thermohaline mixing are hard to make, given the theoretical
uncertainty regarding the efficiency of this
process, but the effect of the FDU can be quantified. This is shown by Fig.~\ref{DNfdu}, which displays the run with age
of $\Delta$[N/Fe], defined as the difference of surface
[N/Fe] between a RGB with P2 composition, and a coeval one with P1 composition (for [Fe/H]=$-$1.3
in this example).

The figure shows both a representative 
initial $\Delta {\rm [N/Fe]_{ini}}$ --the same for all ages-- and the
corresponding differences at the completion of the FDU ($\Delta {\rm [N/Fe]_{FDU}}$).
$\Delta {\rm [N/Fe]_{FDU}}$ is always lower than $\Delta {\rm [N/Fe]_{ini}}$, following
a trend with age, despite the fact that $\Delta {\rm [N/Fe]_{ini}}$ is the same at all ages.
In younger RGB models $\Delta {\rm [N/Fe]_{FDU}}$ is much smaller than $\Delta {\rm [N/Fe]_{ini}}$, getting
progressively closer to its initial value with increasing age.
The reason for this trend is that in RGB models with P1 initial
N abundance, the surface [N/Fe] at the end of the FDU
increases with decreasing age, while the impact of
the FDU is always much reduced or negligible in N-enhanced P2 models. The effect of the FDU on
the surface carbon abundance is also small or negligible in
P2 models with initial C-depleted abundances, while  the surface [C/Fe] at the end of the FDU gets progressively lower 
with decreasing age in P1 models, hence the carbon abundance spread also increases with decreasing age (starting
for a initial constant value at all ages).

\begin{figure}
\centering
\includegraphics[width=8.0cm]{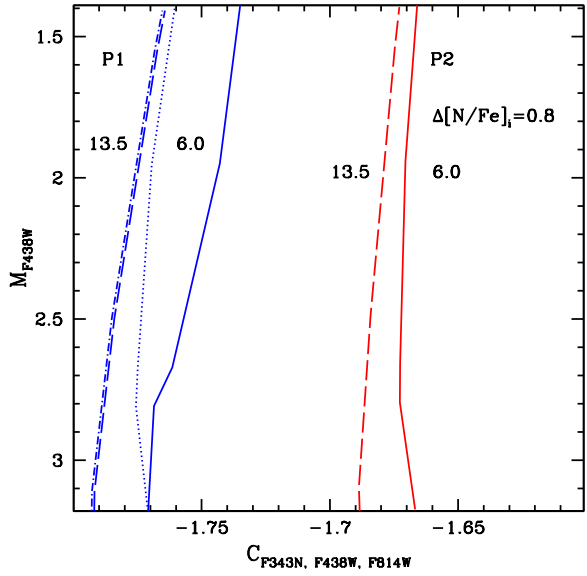}
\caption{$M_{F438W}-C_{F343N,F438W,F814W}$ diagram of P1 and P2 RGB isochrones (calculated for [Fe/H]=$-$1.30)  
          for 6 (solid lines) and 13.5 (dashed lines) Gyr,
          $\Delta {\rm [N/Fe]_{ini}}$=0.8, accounting for the FDU.
          The dotted and dash-dotted lines display
          P1 RGBs calculated without considering the surface abundance variations due to the FDU on the SED,
          for the same initial chemical
          compositions and ages (see text for details).}
\label{fducunbi}
\end{figure}

\begin{figure}
\centering
\includegraphics[width=8.0cm]{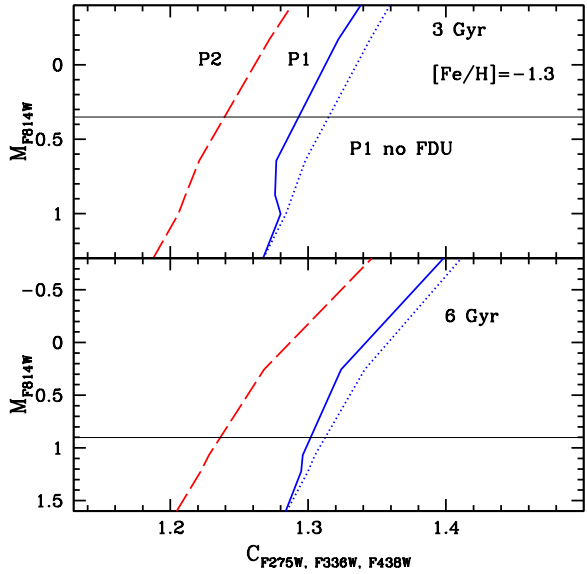}
\caption{$M_{F814W}-C_{F275W, F336W, F438W}$ diagram of 
  P1 (solid lines) and P2 (dashed lines) isochrone RGBs (calculated for [Fe/H]=$-$1.30)
          for ages equal to 3 (upper panel) and 6 (lower panel) Gyr,
          $\Delta {\rm [N/Fe]_{ini}}$=0.8, including the effect of the FDU on the SEDs.
          The dotted lines display P1 RGBs calculated without considering the effect of the
          surface abundance variations due to the FDU.
          The horizontal thin lines mark the brightness corresponding to two magnitudes above the TO,  
          where the width of the RGB is taken in the chromosome maps.}
\label{fduchromo}
\end{figure}

Figures~\ref{fducunbi} and ~\ref{fduchromo} display 
a couple of examples of the impact of the FDU on pseudocolours sensitive to P2 compositions \citep[see][]{s20}.
The P2 chemical composition in this case has [C/Fe]=$-$0.28, [N/Fe]=0.80 and [O/Fe]=$-$0.28
(wich keeps the CNO sum unchanged compared to the P1 metal distribution), whilst the P1 composition is solar scaled;
no variations of Na, Mg and Al have been included in the P2 composition,
because they do not have any relevant effect on the SED.

Figure~\ref{fducunbi} shows theoretical  
RGBs for two pairs of bimodal P1-P2 populations with ages equal to 6 and 13.5~Gyr, [Fe/H]=$-$1.3, 
and $\Delta {\rm [N/Fe]_{ini}}$=0.8, displayed in the $M_{F438W}-C_{F343N,F438W,F814W}$ diagram employed by \citet{mart19}
to study MPs in intermediate-age clusters in the Magellanic Clouds.
The range of $M_{F438W}$ magnitudes corresponds approximately to the range
employed in \citet{mart19} analysis, which encompasses the entire development of the FDU, and stops below the RGB bump.
%The $C_{F343N,F438W,F814W}$ separation between P1 and P2 RGBs at 6 Gyr clearly decreases with
%decreasing magnitude, due to the effect of the FDU.
%The variation (decrease) 
%of C plays a much smaller role, but has the same qualitative effect of the increase of N, that is to shift the
%RGB to larger values of $C_{F343N,F438W,F814W}$.
In the same figure the dotted lines display P1 RGBs that do 
not account for the effect of the FDU on the SEDs.
At 13.5~Gyr the no-FDU (in the SED calculations) RGB
is almost coincident with the FDU RGB, because at this metallicity and age the effect of the FDU
on the surface abundances is very small also for the P1 composition.
For the 6~Gyr case, the no-FDU P1 isochrone RGB runs parallel to the P2 one, and diverges
steadily from the calculations that include the FDU.

Figure~\ref{fduchromo} displays instead the $M_{F814W}-C_{F275W, F336W, F438W}$ RGB diagram (below the RGB bump) of
two P1-P2 bimodal populations (again at a representative [Fe/H]=$-$1.3) with ages of 3 and 6~Gyr,
respectively, including also P1 RGBs calculated
without accounting for the FDU in the SEDs. The $C_{F275W, F336W, F438W}$  N-sensitive
pseudocolour is also employed as the vertical axis of chromosome maps. 
The behaviour is the same as in Fig.~\ref{fducunbi}, just this time the relative position of P1 and P2 RGBs is swapped. 
The level at which the total RGB width is normalized in the chromosome maps to compare different clusters 
(two magnitudes above the main sequence turn off in the $F814W$ filter) is marked.

It is clear how with decreasing age the normalization is at magnitudes increasingly more affected by FDU abundance variations.
Any interpretation in terms of $\Delta {\rm [N/Fe]_{ini}}$ of the RGB widths in the chromosome maps of
intermediate-age clusters, but also comparisons of observed $\Delta_{C\ F275W, F336W, F438W}$ as a function of age, 
must account for the effect of the FDU on the surface [N/Fe] and pseudocolours. 
The net effect of neglecting the influence of the FDU on the model SEDs 
is an underestimate of the initial N-abundance ranges from RGB photometry.

\section{RGB bump luminosity and HB morphology as diagnostics of He abundance spreads}
\label{He}

\begin{figure}
\centering
\includegraphics[width=8.0cm]{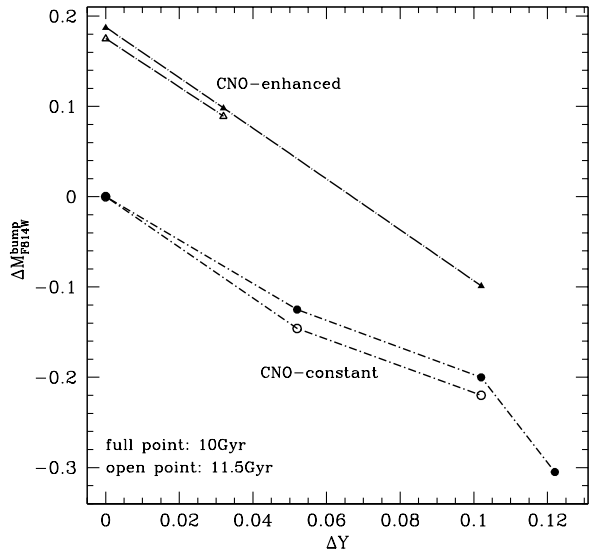}
\caption{Magnitude difference in the ACS $F814W$ filter between the RGB bump brightness of P2 and P1 isochrones (P2 magnitude minus
  P1 magnitude) as a function of the $Y$ difference ($\Delta Y$) between the two populations (P2 $Y$ abundance minus
  the P1 value), for the two labelled ages. We display the case of P2 isochrones with the same CNO sum as the P1 models, and P2
  isochrones with CNO sum enhanced by a factor of 2, as discussed in Sect.~\ref{sec:scenario}. The RGB bump disappears beyond
  threshold values of $Y$, which depend on the isochrone age and chemical composition.}
\label{bumpy}
\end{figure}

As mentioned in Sect.~\ref{sec:scenario}, the luminosity of the RGB bump at fixed $Z$ and age
is sensitive to the initial $Y$, hence it can potentially
be used to constrain the relative helium abundances among cluster MPs, if 
high precision photometry is available.
As an example, Fig.~\ref{bumpy} displays the predicted 
magnitude difference in the ACS $F814W$ filter (insensitive to light element abundance anticorrelations)
between the RGB bump brightness of P2 and P1 $Z$=0.002 isochrones, as a function
of the difference in $Y$ ($\Delta Y$, calculated with respect to the value $Y$=0.248 for the P1 composition),
for two ages and P2 compositions with constant (the same as P1 isochrones)
and enhanced (by a factor of 2) CNO, respectively.
For a given cluster age and metallicity, measurements of magnitude differences of the bump brightness translate 
straightforwardly into values of $\Delta Y$.

The dependence on age increases with increasing $\Delta Y$, whilst the relationship for CNO-enhanced P2 isochrones
is offset by a sizeable zero point, due to the fact that for the same $Y$ and age the RGB bump in CNO enhanced isochrones
is fainter than in $\alpha$-enhanced P1 ones. 
It is also important to notice that the RGB bump disappears 
beyond threshold values of $Y$, which depend on age and chemical composition. 

\citet{bragaglia:10} determined empirically that the $V$ bump luminosity for P2 stars (identified spectroscopically)
in a sample of 14 Galactic GCs is on average brighter than P1 stars, implying larger $Y$ abundances.
\citet{nataf:11b} concluded that the gradient of the bump brightness and star
counts with the radial distance in the Galactic GC 47~Tuc, is  
consistent with the presence of a helium enriched stellar population
in the cluster centre.
More recently, \citet{milone:15} detected the RGB bump in three
populations along the RGB of
of NGC2808, as identified from the chromosome map, deriving helium differences $\Delta{Y} \sim$ 0.03
and $\sim$0.10, consistent with results obtained from the MS colours (see Sect.~\ref{Heb}).

\citet{lagioia} have determined P1 and P2 RGB bump magnitudes
for a sample of 26 Galactic GCs, using chromosome maps to separate the two types of populations.
For 18 clusters they were able to determine from optical CMDs an average 
$\Delta Y$ = 0.011$\pm$0.002 between P1 and P2 stars (assuming coeval populations), the latter being more helium rich.
\citet{lagioiab} performed a similar analysis in 4 Small Magellanic Cloud old clusters, finding for three of them
a slightly larger average $Y$ in P2 stars.

Also the HB can be used to estimate He abundance spreads in intermediate-age \citep{chante} and old star clusters
\citep[see, e.g.,][]{dantona:02, dicri, dale11, dale13, gratton:13, nieder17}, because its morphology is very sensitive
to the initial He distribution of the cluster population, as shown by the next two figures.

\begin{figure}
\centering
\includegraphics[width=10.0cm]{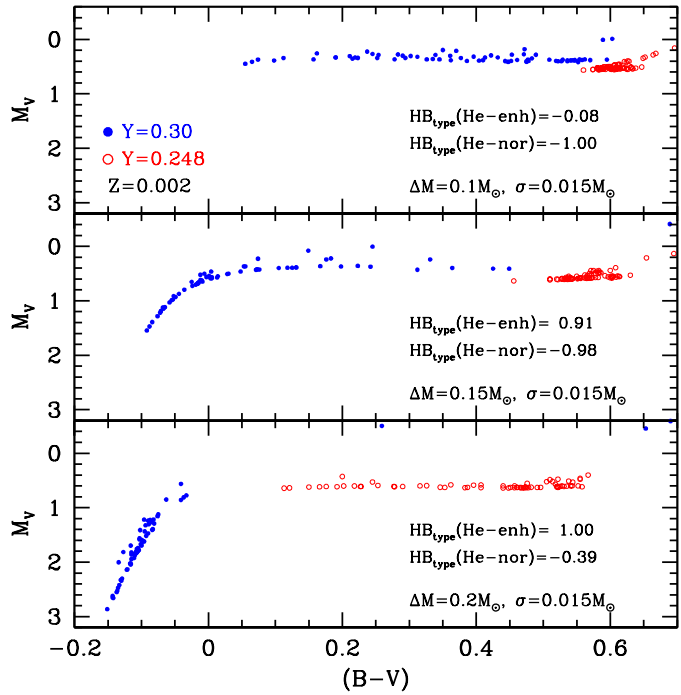}
\caption{Three optical CMDs of a synthetic HB for a bimodal P1-P2 cluster with
  the labelled $Z$ and the two labelled values of $Y$. The RGB progenitor masses correspond to
  an age of 12.5~Gyr. From top to bottom, the mean value $\Delta M$ of the
  mass lost by the RGB progenitors (assumed to be the same 
  for both P1 and P2 objects) increases. The assumed 1$\sigma$ Gaussian spread around $\Delta M$ is also labelled.
  The value of the morphology parameter $HB_{type}$ for both P1 and P2 objects
  is displayed in each panel (see text for details).}
\label{hbsynth}
\end{figure}

Figure~\ref{hbsynth} displays optical CMDs (hence BCs are not affected by the
light element abundance variations) of three synthetic HBs,
each one calculated for a bimodal P1-P2 cluster with uniform
metallicity ($Z$=0.002), the two labelled initial values of $Y$ ($Y$=0.248 for P1 stars and $Y$=0.300 for P2 stars), 
and RGB progenitor ages equal to 12.5~Gyr, corresponding to initial masses equal to 0.84$M_\odot$ and 0.76$M_\odot$ for
the P1 and P2 population, respectively.
The mean amount of mass lost by the RGB progenitors ($\Delta M$) 
is the same for both populations in each CMD, and increases from
the top to the bottom panel, as labelled. The spread around these mean values is the same in
all diagrams.

Regardless of the specific value of $\Delta M$, assuming it is the same for both P1 and P2, 
the He-enhanced P2 synthetic objects are bluer than the P1 ones. This a 
consequence of the lower RGB progenitor mass in the He-enhanced population.
With increasing $\Delta M$, the overall HB morphology becomes bluer, and a blue tail typical of optical HB CMDs (due to
the large increase of the BCs above a threshold $T_{\mathrm eff}$) start to appear.
A quantitative measure of this increasingly bluer morphology with increasing $\Delta M$ is provided by the 
$HB_{type}$ parameter, defined as $HB_{type}\equiv {\rm (B-R)/(B+V+R)}$ where B, R, and V are
the number of objects bluer, redder and within the RR Lyrae instability strip, respectively. The values
of $HB_{type}$ for both P1 and P2 objects increase with increasing $\Delta M$, as expected from the progressive shift
to bluer colours of the synthetic populations.

Due to the lack of a reliable theory for the RGB mass loss, we cannot be sure that $\Delta M$ is truly the same
in P1 and P2 stars, and in terms of colour distributions, morphologies like the ones in Fig.~\ref{hbsynth} could
be reproduced by just P1 objects, with a bimodal distribution of $\Delta M$ values.
However, the enhancement of He affects also the magnitudes --apart from the objects on the
blue tail in this optical CMD, where He-normal and He-enhanced objects are located essentially on the same sequence.
The P2 He-enhanced population is brighter, because of the increased efficiency of the H-burning shell
(see Sect.~\ref{sec:scenario}), and this increase of luminosity cannot be reproduced by tuning the mass loss of P1
objects.

\begin{figure}
\centering
\includegraphics[width=8.0cm]{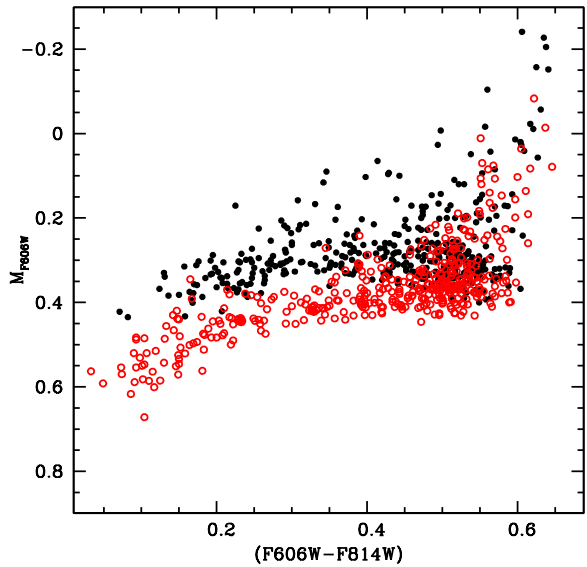}
\caption{Optical CMD in the ACS filter system of two synthetic HBs with the same metallicity, progenitor age,
  but in one case (open circles) a constant $Y$=0.248 and a large spread of $\Delta M$, whilst in the other
  simulation $\Delta M$ has a negligible spread but a range of helium abundances $\Delta Y$=0.04
  (filled circles). Both simulations cover
  approximately the same colour range (see text for details).}
\label{hbsynthb}
\end{figure}

Figure~\ref{hbsynthb} makes this point very clear, by showing the 
optical CMD in ACS optical filters, of two synthetic HBs with the same metallicity and progenitor age,
  but in one case a constant initial $Y$=0.248 and a large spread of $\Delta M$ (with uniform distribution), whilst in the other
  simulation $\Delta M$ has a single value with a negligible spread,
  but a range of initial helium abundance enhancements (uniform distribution),
  from 0 to $\Delta Y$=0.04. Both simulations cover
  approximately the same colour range, but the slope of the synthetic HB is clearly different in the two cases. A 
  different slope of the HB when comparing synthetic populations with constant He and large spread of $\Delta M$,
  with populations at constant $\Delta M$ and large spread of $Y$, appears also in simulations of red HBs like the Galactic GC
  47~Tuc, and intermediate age clusters \citep[see, e.g.][]{scp, chante}.

  A range of initial He abundances explains the 
very tilted optical CMDs of the HB in Galactic GCs like NGC~6388, NGC~6441 and
NGC~1851\citep[see][and references therein]{caloi:07, busso:07, scp:08, tailo:17}, and also 
extended blue tails like in the CMDs of NGC~2808 \citep[]{dantona:05, lee:05}, and again NGC6441 and NGC6388.

Regarding the blue tails in UV CMDs, \citet{dale11} and \citet{dale13} have shown with extensive
simulations how  the use of UV filters for these hot HB stars help to disentangle the degeneracy between
He-normal and He-enhanced populations, by properly tracing the differences in bolometric luminosities
discussed in Sect.~\ref{sec:scenario}.

Finally, the expectation that He-rich stars populate preferentially the hotter side of the HB in old clusters,
has been used to infer the presence of MPs from the integrated colours of extragalactic globular clusters.
\citet{bellini} have employed integrated UV-optical CMDs and two-colour diagrams to identify a small group of likely GCs 
belonging to the elliptical galaxy M87, which might host multiple populations with extreme helium content.

\section{The main sequence as tracer of He abundance spreads}
\label{Heb}

\begin{figure}
\centering
\includegraphics[width=8.0cm]{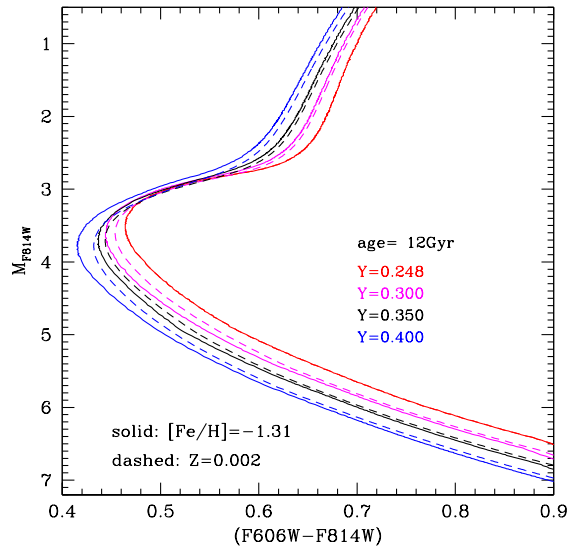}
\caption{Optical CMD from the MS to the lower RGB
  in the ACS filter system, for 12~Gyr isochrones with the various labelled initial values of $Y$ and either constant [Fe/H] or
  constant $Z$.}
\label{figcmdy}
\end{figure}

As discussed in Sect.~\ref{SED}, spreads and/or splittings of the MS in optical CMDs provide us with yet another way 
to determine He-abundance spreads in clusters, because they 
are caused by variations of the initial $Y$ between P1 and P2 stars, being independent of associated
light-element abundance variations. Figure~\ref{figcmdy} shows representative coeval isochrones (with [$\alpha$/Fe]=0.4)
in an optical CMD, calculated by varying the initial $Y$ at either fixed $Z$ or fixed [Fe/H].
The behaviour mirrors what was discussed in the theoretical HRD.
Using the MS in optical colours like $(F606W-F814W)$ to detect He abundance spreads in clusters 
has the additional advantage of a weak metallicity dependence of the $(F606W-F814W)$-$T_{\mathrm eff}$ relation \citep{bastiss}.

\begin{figure}
\centering
\includegraphics[width=8.0cm]{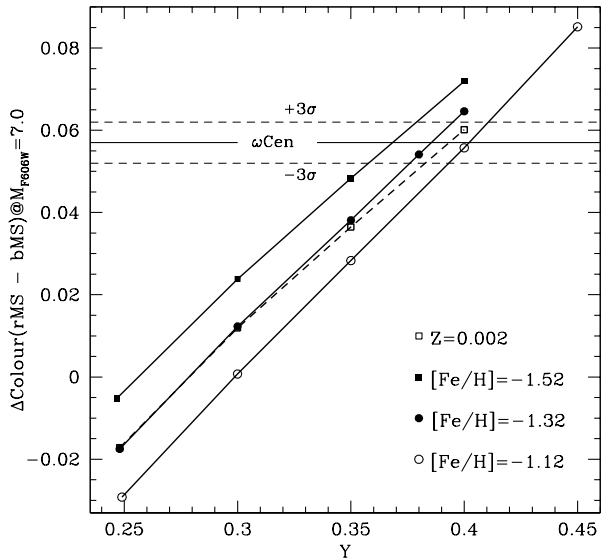}
\caption{Difference of the MS colour $(F606W-F814W)$ ($\Delta{(F606W-F814W)}$)
  as a function of the initial $Y$ taken at $M_{F606W}$=7, 
  for sets of isochrones with the labelled [Fe/H] values (constant [Fe/H] at varying $Y$ -- solid lines). 
  The [Fe/H] values are consistent with spectroscopc measurements along the blue and red MS of $\omega$~Cen, and 
  the horizontal solid line shows the value of $\Delta (F606W-F814W)$ between the red and blue MS as measured in this
  cluster by 
  \cite{king:12}. The horizontal dashed lines represent the $3\sigma$ uncertainty of this measurement. The long-dashed line shows
  the predicted difference when keeping fixed $Z$ when varying $Y$ (see the text for more details).}
\label{figdycol}
\end{figure}

The determination of He abundance spreads from the MS is straighforward: It is 
enough to evaluate colour differences on the MS at a given magnitude level, for example
$\sim2$~mag below the TO to avoid age effects \citep{cassisi:17}, and compare them with model predictions.
This is outlined in Fig.~\ref{figdycol}, as applied to the  case of the blue and red sequences 
observed along the MS of $\omega$~Cen \citep{bedin:04,piotto:05,king:12}, a more complicated situation 
than for standard monometallicity GCs. The figure displays the difference of the
$(F606W-F814W)$ colour taken 
at $M_{F606W}=7.0$ between an isochrone with the same [Fe/H] of the red MS in the cluster ([Fe/H]=$-$1.62,
assuming it has standard $Y$=0.246) and an isochrone with
the [Fe/H] of the blue MS (the three possible values labelled),  
as a function of the initial He abundance of the isochrone representative of the blue MS. 
The [Fe/H] of the blue MS isochrones
is kept constant when $Y$ varies, fixed at the measured value. The three different [Fe/H] for the blue MS
reflect the observational uncertainty \cite[see, e.g.,][]{king:12}, and 
the measured value of the colour difference is also marked, together with the $\pm 3 \sigma$ uncertainty.
The fact that the red MS is more metal poor than the blue one means that the blue MS must have a much higher initial $Y$.

%In any case, we have to keep in mind
%that in this methodological approach is the difference of the metallicities that matters, rather than the absolute value of either
%of them: in case one would shift both the bMS and rMS [Fe/H] values by 0.1Ð0.2 dex or so, there would not be any appreciable effect
%on the helium abundance derived for the bMS. The horizontal line shows the empirically obtained value, while the two horizontal
%dashed lines correspond to the empirical measurement but accounting also for a $\pm3\sigma$ empirical uncertainty in the measurement.

The intersection of the solid horizontal line with the predictions from the theoretical isochrones at varying $Y$ 
provides a direct estimate of $Y$ for the blue MS.
Values derived considering the full range of estimates of [Fe/H] for the blue MS provide 
an estimate of the effect of this source of error, to combine with error from the uncertainty on the measured value
of the colour difference.

%It has been previously discussed the difference in the location of He-enhanced isochrones when they are computed by keeping constant
%either the global metallicity $Z$ or [Fe/H] (see also Fig.~\ref{figcmdy}).
Figure~\ref{figdycol} shows also a prediction for the colour difference between the blue and red MS of $\omega$~Cen 
by keeping fixed the metallicity $Z$ of the blue MS when increasing $Y$.
The assumed value $Z$=0.002 corresponds exactly to [Fe/H]=$-1.32$ (with a [$\alpha$/Fe]=0.4 metal distribution)
for a {\sl normal} $Y$=0.248, but to keep [Fe/H] fixed to this value with increasing $Y$, the metallicity $Z$ should increase by
for example $\sim$15\% for $Y$=0.40. This explains the difference with the results for [Fe/H]=$-$1.32, constant with $Y$.
There is no significant variation in the colour difference predicted by the $Z$ constant model until $\Delta{Y}\approx0.05$.
Above this limit, the constant $Z$ models predict systematically larger $Y$ for the blue MS.

For the bulk of globular clusters the interpretation of colour differences at fixed magnitude
along the MS due to MPs with different $Y$ is much simpler.
The measured values would correspond to differences calculated from 
isochrones with the same $Z$ and varying $Y$ (because of the uniform Fe abundance in the cluster).
The minimum value of the difference
would then be zero, when both points belong to an isochrone with the same $Z$ and $Y$.
On the contrary, in case of $\omega$~Cen, when $Y$ of the blue MS is the same as for the red one, the predicted colour difference
is negative, meaning that the blue MS should actually be redder than the red one, because of its higher [Fe/H].

The use of optical CMDs of the cluster MS as He abundance diagnostic has been discussed in detail by
\cite{cassisi:17} who found from their $\alpha$-enhanced models that:

\begin{itemize}

\item The derivative $dY/d(F606W-F814W)$ (taken two magnitudes below the TO in $F606W$)
  is weakly dependent on the isochrone metallicity, 
%  as also shown in Fig.~\ref{figdycol};
  being equal to about $2.3~mag^{-1}$ at $Z$=0.0003 (corresponding to [Fe/H]=$-2.1$ for $Y$=0.245)
  and to 1.7~$mag^{-1}$ at  $Z$=0.008 (corresponding to [Fe/H]$=-0.7$ for $Y$=0.256);

\item The values of $dY/d(F606W-F814W)$ are also largely independent of the exact value of the
  MS reference brightness. A change from 2 magnitudes to 1 magnitude below the TO at $Z$=0.002 (corresponding to [Fe/H]=$-$1.3 
  for $Y$=0.248) changes $dY/d(F606W-F814W)$ from 1.8~$mag^{-1}$ to 2.1~$mag^{-1}$;

\item $dY/d(F606W-F814W)$ is also fairly independent of the isochrone age when the reference brightness
   is at least about 2~mag below the TO, in the unevolved part of the MS.
\end{itemize}

In the following, we discuss briefly the accuracy of the He abundance spreads derived from MS optical CMDs,
following \citet{cassisi:17}. We summarize the impact of the efficiency of superadiabatic convection, the
 treatment of the outer boundary conditions of the models, the effect of atomic diffusion, and 
 the choice of BCs. The tests discussed below are all made at fixed $Z$=0.002, [$\alpha$/Fe]=0.4,
 corresponding to [Fe/H]=$-$1.32 for $Y$=0.248.
% appropriate for MPs in metal-intermediate Galactic GCs.
 
\begin{figure}
\centering
\includegraphics[width=8.0cm]{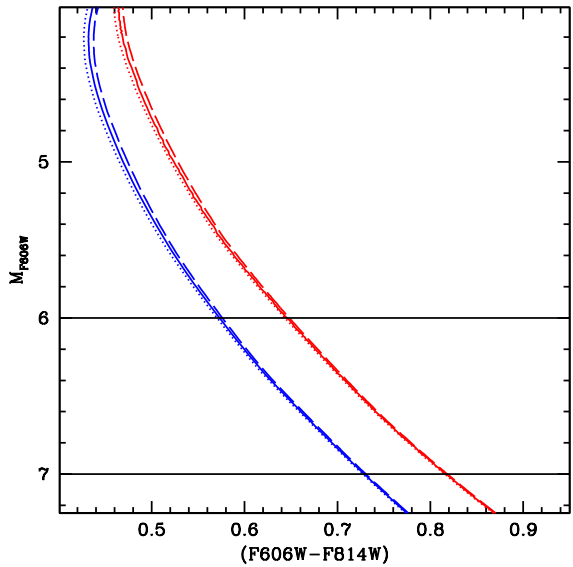}
%%\vskip -0.5cm
\caption{Optical CMD of 12~Gyr isochrones for $Z=0.002$, $Y=0.248$ and 0.40, computed with a solar value
  of $\alpha_{\rm MLT}$ (solid lines) or alternatively increasing (dotted lines) and decreasing (dashed lines) the solar 
  value by 0.1. The magnitude levels corresponding to $M_{F606W}$=6 and 7 (about 2 and 3 magnitudes below the TO) are marked.}
\label{isomsmlt}
\end{figure}

Regarding the treatment of the superadiabatic envelope convection, the mixing length theory
\citep[MLT --][]{bv} is almost universally used.
This formalism contains in its standard form 4 free parameters: three parameters 
are fixed a priori (and define what we denote as the MLT {\sl flavour}) 
whereas one (the so-called mixing length, $\alpha_{\rm MLT}$) is calibrated by 
reproducing observational constraints, firstly the solar radius by means of a solar model.
Tracks computed with different MLT flavours do basically overlap
along MS, SGB and RGB as long as $\alpha_{\rm MLT}$ is appropriately calibrated to match the solar radius 
\citep[see, e.g.,][]{mlt,salaris:18}.
However, given that there is no {\sl a priori} reason for the solar $\alpha_{\rm MLT}$ 
to be appropriate also for other masses, chemical compositions and evolutionary phases, 
it is important to check how $dY/d(F606W-F814W)$
is affected by changes of $\alpha_{\rm MLT}$.
Figure~\ref{isomsmlt} shows how the CMD location of the MS for two values of $Y$ changes when 
$\alpha_{\rm MLT}$ is varied by $\pm0.1$ around the solar value.
This variation is consistent with the changes along the MS predicted by the 3D radiation hydrodynamics 
models by \cite{tramp:14} and \cite{magic:15}.
At a given He abundance the three  sequences are almost coincident for
$M_{F606W}\sim$6 (about 2 magnitudes below the TO) as well as at fainter magnitudes. The reason is that models in this magnitude
range have deep convective envelopes
almost completely adiabatic, and the variation of $\alpha_{\rm MLT}$ has a negligible effect on their temperature stratification, 
hence on the predicted $T_{\mathrm eff}$ and colours.
%In addition, since there is no reason to assume different $\alpha_{\rm MLT}$ values
%for the He-rich and He-poor sub-populations,  the calibration of d$Y$/d($F606W-F814W$) is completely unaffected by the MLT
%calibration issue.

To integrate the stellar structure equations, it is necessary to fix the value of the pressure and temperature at the photosphere
  \citep[see, e.g.,][and references therein]{don:08}. The choice of these outer boundary conditions has
an impact of the MS location, because it affects the $T_{\mathrm eff}$ of stellar models with convective envelopes
\citep[see, e.g.,][for more details]{cs:13}. As a test,
we have calculated isochrones computed adopting two widely used $T(\tau)$ relationship
to determine the model boundary conditions, namely the \cite{ks} $T(\tau)$ and the Eddington grey $T(\tau)$. For each
of these two choices 
the value of $\alpha_{\rm MLT}$ is calibrated on the Sun.
Calculations with the Eddington $T(\tau)$ are systematically
bluer by $\approx0.02$~mag, but the shift is independent of the adopted initial He abundance, the shape of the MS
is preserved, and the value of $dY/d(F606W-F814W)$ is unchanged.

%\begin{figure}
%\centering
%\includegraphics[width=8.0cm]{figisohebound.eps}
%%\vskip -0.5cm
%\caption{As Fig.~\ref{isomsmlt} but showing 
%  isochrones calculated using either the \cite{ks} $T(\tau)$ relation to determine the model outer boundary conditions
%  (solid lines), or the Eddington grey $T(\tau)$ (dashed lines).}
%\label{isomsbound}
%\end{figure}
 
\begin{figure}
\centering
\includegraphics[width=8cm]{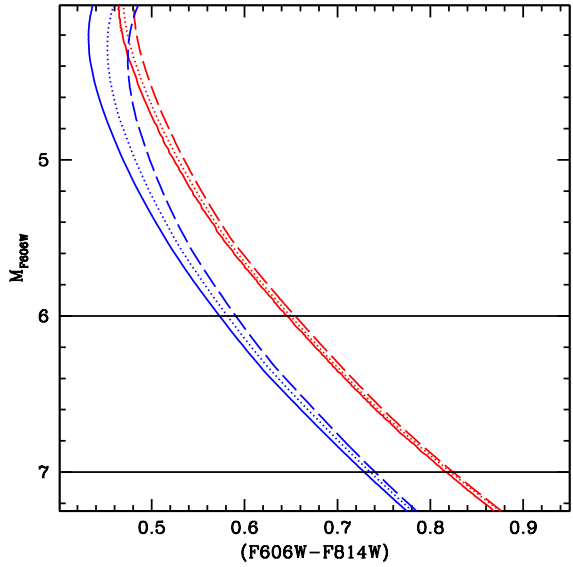}
%%\vskip -0.5cm
\caption{As Fig.~\ref{isomsmlt} but displaying isochrones computed from models calculated without atomic diffusion
  (solid line), fully efficient diffusion throughout the models, like in the solar model (dashed line),
  diffusion from the convective envelope inhibited, but efficient in the radiative interior (dotted line).
  For the cases where atomic diffusion was included in the calculations, the age of
  the isochrone has been reduced by 1~Gyr to account for the reduction of the evolutionary lifetimes, 
  so that all isochrones have a very similar TO brightness.}
\label{isomsdif}
\end{figure}

It is well known \citep[see, e.g.][]{cs:13} that, although helioseismic data require the inclusion of
approximately fully efficient atomic diffusion in the standard solar model, 
 high resolution spectroscopy of metal-poor, Galactic GC stars has shown strong evidence that the efficiency 
 of this element transport mechanism has to be lower than predicted by theory.
 Atomic diffusion affects the chemical stratification of the MS models,
 modifying opacity, and impacting the model $T_{\mathrm eff}$ scale.
 Following the detailed investigation by \cite{cassisi:17}, Fig.~\ref{isomsdif} compares isochrones
 with no atomic diffusion, fully efficient diffusion thorughout the models, and
 diffusion from the convective envelope inhibited, but efficient in the radiative interior.
 The first two cases correspond to the two extreme 
situations: Fully efficient atomic diffusion shifts the isochrone MS to redder colours, by an amount that increases moving
towards the MS turn-off. However, the faintest portion of the MS is less affected, because models
have very extended (in mass) fully mixed, convective envelopes, which minimize the 
surface depletion of metals and He due to atomic diffusion. This has the important implication that the calibration of
$dY/d(F606W-F814W)$  depends in this case on the adopted reference magnitude: At $M_{F606W}$=7, the effect of
diffusion is barely noticeable, whilst it is more relevant at $M_{F606W}$=6. 

The impact on the MS $T_{\mathrm eff}$ and colours is He dependent, increasing with increasing initial He abundance,
because the larger $Y$, the shallower the convective envelope. When diffusion is restricted to the radiative interiors
the impact on the MS colours is obviously reduced.
In case of fully efficient diffusion $dY/d(F606W-F814W)$ at $M_{F606W}$=6 increases 
by about 0.4~$mag^{-1}$ in the He range between 0.30 and 0.35,
and by about 0.5~$mag^{-1}$ for initial He abundances larger than 0.35,
compared to no diffusion models.

\begin{figure}
\centering
\includegraphics[width=8cm]{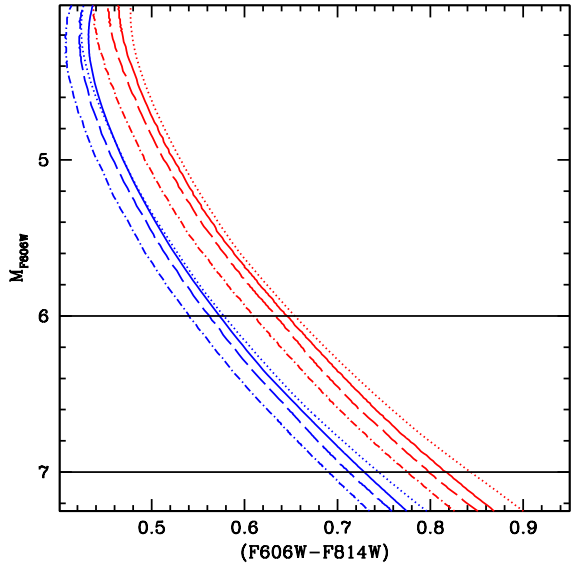}
%%\vskip -0.5cm
\caption{As Fig.~\ref{isomsmlt}, but showing isochrones for four different choices of BCs. 
  Solid lines show the case of ATLAS9 bolometric corrections \citep[]{bastiss}, dashed lines PHOENIX BCs \citep[]{bh},
  dash-dotted lines MARCS BCs \citep[]{cv}, whilst dotted lines correspond to the empirical BCs by \cite{wl}.}
\label{isobc}
\end{figure}

Finally, we briefly discuss the choice of BCs. Figure~\ref{isobc} shows the comparison of isochrones for two different
values of $Y$ ($Y$=0.248 and $Y$=0.40) and four independent choices of BCs.
Three sets come from theoretical calculations \citep[ATLAS9, employed as a reference in all previous tests,
  PHOENIX, MARCS, from][respectively]{bastiss, bh, cv}, whilst one set is empirical \citep{wl}.
Varying the set of BCs affects the colours of the MS, but the impact on $dY/d(F606W-F814W)$ is minimal when considering
the three theoretical BC calculations, because differences are just systematic shifts, the same for all $Y$.
On the other hand, the use of the empirical BCs does 
impact $dY/d(F606W-F814W)$, which changes compared to the reference values for the ATLAS9 BC, because of a clear change
of shape of the isochrones.
%there is a shift in the MS colours (generally redder than the reference isochrones) but also a clear change of shape.
The value of $dY/d(F606W-F814W)$ is almost the same as for the reference ATLAS9 case at $M_{F606W}$=6, but it gets
increasingly smaller than the corresponding reference value when considering increasingly fainter magnitudes along the MS.

\section{Present status and outlook}
\label{concl}

The previous sections have discussed in some detail a range of photometric methods, informed by the results of
high resolution spectroscopy, to disentangle and characterize the main properties of MPs in massive star clusters.
Their application to Milky Way and Magellanic Clouds' clusters has had a massive impact on studies in the field.
Below, we give a concise summary of the main information gathered by photometric investigations of MPs
\citep[see, e.g.,][and references therein]{dale14, m17, bl18, He18, lsb18, lagioiab, lagioia19, chante, martocchia18, mart19},
which set strong constraints for 
all scenarios/models put forward to explain
their origin, and more in general the formation of massive star clusters.

\begin{itemize}

\item Figure ~\ref{agemass} summarizes, in a cluster (present-day) mass-age diagram,
  the situation regarding photometrically detected MPs
in Galactic GCs and massive clusters of the Magellanic Clouds.
MPs have been detected down to ages around 2~Gyr, but not below this threshold.
At ages below about 8 Gyr the lower (present-day) mass limit for the presence of MPs
is around $10^5 M_{\odot}$, but several older clusters with MPs have lower masses,
down to about $2 \times 10^4  M_{\odot}$.

\begin{figure}
\centering
\includegraphics[width=10.0cm]{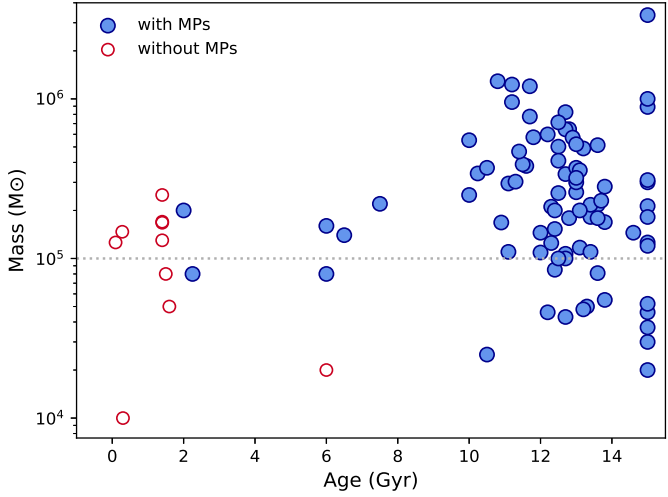}
\caption{Present-day mass vs. age diagram of surveyed Galactic GCs and massive clusters in the
  Magellanic Clouds,
  with and without photometrically detected MPs. A reference mass equal to $10^5 M_{\odot}$ is marked (see text for details).}
\label{agemass}
\end{figure}

\item In some clusters clear multimodal (pseudo-)colour distributions are detected in RGB chromosome maps or CMDs, whilst
  in other cases smooth distributions are observed, without evident separations between populations.

\item P2 stars are generally the most abundant population in most Galactic GCs.
  On average, the larger the cluster mass, the smaller the fraction of P1 stars.

\item A substantial fraction of GCs display internal chemical variations in P1 stars,
  very likely associated to a range of initial He abundances or, possibly, small spreads of Fe on the order of 0.1~dex.

\item In Galactic GCs the RGB width in colours and pseudocolours sensitive to He and N abundances
  correlates strongly with the cluster metallicity. When this dependence is removed, 
  the RGB width correlates with cluster mass and luminosity. More massive and brighter clusters have larger
  internal variations of these elements. When considering also the relationship between number fraction of P2 stars and
  cluster mass, it turns out that the larger the fraction of P2 stars, the larger the internal variations of He and N.

\item Most GCs have P2 stars more concentrated in the innermost region than P1 stars, with few exceptions whereby
P1 stars are more centrally concentrated than P2 stars, or both P1 and P2 stars share the same radial distribution.

\item The current sample of observed Magellanic Clouds' clusters display a trend with age of the RGB width
  in nitrogen sensitive indices. 
  Even after correcting for the effect of FDU, the width of the RGB increases with
  increasing age. The range of He abundances derived from the CMD of the core He-burning phase
  however does not seem to correlate with age, and follows the same trend with cluster mass found in Galactic GCs.

\item Estimates of the age difference between P1 and P2 populations (when they have been determined) in Galactic 
  GCs are within $\sim$300~Myr \citep[see, e.g.,][]{marino:12, nardiello15}, and within $\sim$20~Myr for massive clusters
  about 2~Gyr old \citep[see, e.g.,][]{martocchia18b, sara20}.

\end{itemize}

The discovery of MPs in intermediate-age massive clusters of the
Magellanic Clouds --the same MP phenomenon as in Galactic GCs-- has certainly
put into a new perspective the problem of their formation.
Whilst it has been generally acknowledged that MP formation was connected to high-redshift environments, their discovery in
clusters with ages as low as 2~Gyr is clearly in serious conflict with this idea.
An increase of the sample of Magellanic Clouds' clusters hosting MPs (at the moment we have about just 
10 clusters with MP detection)
is certainly necessary, together with a homogeneous comparison with Galactic GCs using consistent pseudocolours
\citep[see][]{lagioia19} and chromosome maps \citep[see][]{sara19, sara20}.

In the near future JWST will provide us with photometric diagnostics to search for MPs
in resolved massive clusters even beyond the Local Group \citep{s20}. This will allow us to investigate
in more depth the role that the environment might play in the formation of MPs.
At the same time JWST will also enable to study the lower MS in a much larger sample of Galactic GCs than possible today with
HST, and provide crucial information on the present-day mass functions of P1 and P2 stars, plus estimates of 
N and O abundance ranges in this fully convective objects still on the zero age MS,
to compare with results from the much more evolved RGB stars in the same cluster.

\begin{acknowledgements}

 We acknowledge the anonymous referee for his/her helpful suggestions.
  We warmly thank Adriano Pietrinferni for interesting discussions and the fruitful collaboration
  over all these years.
We wish to acknowledge Giampaolo Piotto for his leading role
in the development of photometric studies of MPs in globular clusters.
Raffaele Gratton and Alvio Renzini are also warmly acknowledged for interesting discussions and collaborations on 
this research topic.
We thank Nate Bastian and Carmela Lardo for several comments on an early draft of the manuscript.  
We are grateful to Carmela Lardo also for producing some of the figures.
SC acknowledges support from Premiale INAF MITiC, from Istituto Nazionale di Fisica Nucleare (INFN)
(Iniziativa specifica TAsP), progetto INAF Mainstream (PI: S. Cassisi), PLATO ASI-INAF agreement n.2015-019-R.1-2018, 
and grant AYA2013- 42781P from the Ministry of Economy and Competitiveness of Spain. 

\end{acknowledgements}

\section*{Conflict of interest}
 The authors declare that they have no conflict of interest.

\bibliographystyle{spbasic}      % basic style, author-year citations
\bibliography{mpteo_review.bib}   % name your BibTeX data base

\end{document}